\documentclass[a4paper,11pt]{article}
\pdfoutput=1 

\usepackage{jheppub} 
\usepackage[bottom]{footmisc}
\usepackage{amssymb}
\usepackage{amsmath}
\usepackage{amsthm}
\usepackage[usenames,dvipsnames]{xcolor}
\usepackage{epsfig}
\usepackage{dcolumn}
\usepackage{tikz}
\usetikzlibrary{shapes.geometric, arrows}
\usepackage{upgreek}
\usepackage{setspace}
\usepackage{enumitem}
\usepackage{array,multirow,bigdelim,arydshln}
\usepackage{appendix}
\usepackage{xparse}
\usepackage{stmaryrd}
\usepackage[T1]{fontenc} 
\usepackage{mathtools}
\usepackage{physics} 
\usepackage{adjustbox}
\usepackage{multirow}
\usepackage{graphicx} 
\usepackage{float} 
\graphicspath{{./images/}}
\usepackage[nottoc]{tocbibind}
\usepackage{hyperref}
\usepackage[utf8]{inputenc}
\hypersetup{
	colorlinks,
	urlcolor=Maroon,
	linkcolor=Maroon,
	citecolor=Maroon
	}

\NewDocumentCommand{\binomial}{omm}
 {%
  \genfrac(){0pt}{}{#2}{#3}%
  \IfValueT{#1}{_{\!#1}}%
 }
\NewDocumentCommand{\eulerian}{omm}
 {%
  \genfrac<>{0pt}{}{#2}{#3}%
  \IfValueT{#1}{_{\!#1}}%
 }

\def \s {\sigma}

\usepackage{latexsym}
\usepackage{tikz}

\theoremstyle{plain}

\theoremstyle{definition}

\newcommand{\dedge}{\,\raisebox{2.5pt}{\tikz{\filldraw (0,0) circle(1pt);
\filldraw (.4,0) circle(1pt);
\draw (.4,0)--(0,0);
}}\,}

\def\bea#1\eea{\begin{eqnarray}#1\end{eqnarray}}
\def\be#1\ee{\begin{equation}#1\end{equation}}
\def\ba#1\ea{\begin{align}#1\end{align}}

\def\nl{\nonumber\\}

\def\yz#1\yz {{\color{blue} [[YZ: #1]] }}
\def\hou#1\hou {{\color{blue} [[Hou: #1]] }}
\def\tian#1\tian {{\color{blue} [[Tian: #1]] }}
\newcommand{\dif}{\mathrm{d}}

\usepackage{amsmath}
\usepackage{multicol}
\usepackage{bbm}
\usepackage{enumerate}

\usepackage{amsthm}
\usepackage{mathrsfs}
\usepackage{upgreek}
\usepackage{amssymb}
\usepackage{bm}
\usepackage{setspace}
\usepackage{array,multirow,arydshln}
\usepackage{bigdelim}
\usepackage{scalerel}
\usepackage{diagbox}
\usepackage{caption}
\usepackage{tabularx}

\usepackage{tikz}

\usetikzlibrary{shapes.geometric,arrows,arrows.meta,decorations.pathmorphing,decorations.markings,patterns}

\def\<{\langle}
\def\>{\rangle}

\def\b{\beta}

\def\e{\epsilon}

\def\s{\sigma}

\newcommand*{\halfwayb}{0.5*\pgfdecoratedpathlength+2.4pt}

\usepackage[percent]{overpic}
\usepackage{multirow} 
\usepackage{slashed}

\title{Kinematic numerators from the worldsheet: cubic trees from labelled trees}
\author[a,b,c,d,e]{Song He}\emailAdd{songhe@itp.ac.cn}
\author[a,b]{Linghui Hou}\emailAdd{houlinghui@itp.ac.cn}
\author[b,d]{Jintian Tian}\emailAdd{tianjintian@itp.ac.cn}
\author[f,b]{and Yong Zhang}\emailAdd{yong.zhang@physics.uu.se}

\affiliation[a]{School of Fundamental Physics and Mathematical Sciences, Hangzhou Institute for Advanced Study, UCAS, Hangzhou 310024, China}
\affiliation[b]{CAS Key Laboratory of Theoretical Physics, Institute of Theoretical Physics, Chinese Academy of Sciences, Beijing 100190, China}
\affiliation[c]{ICTP-AP
International Centre for Theoretical Physics Asia-Pacific, Beijing/Hangzhou, China}
\affiliation[d]{School of Physical Sciences, University of Chinese Academy of Sciences, No.19A Yuquan Road, Beijing 100049, China}
\affiliation[e]{Peng Huanwu Center for Fundamental Theory, Hefei, Anhui 230026, P. R. China}
\affiliation[f]{Department of Physics and Astronomy, Uppsala University, 75108 Uppsala, Sweden}

\abstract{In this note we revisit the problem of explicitly computing tree-level scattering amplitudes in various theories in any dimension from worldsheet formulas. The latter are  known to produce cubic-tree expansion of tree amplitudes with kinematic numerators automatically satisfying Jacobi-identities, once any half-integrand on the worldsheet is reduced to logarithmic functions. We review a natural class of worldsheet functions called ``Cayley functions'', which are in one-to-one correspondence with labelled trees, and natural expansions of known half-integrands onto them with coefficients that are particularly compact building blocks of kinematic numerators. We present a general formula expressing kinematic numerators of all cubic trees as linear combinations of coefficients of labelled trees, which satisfy Jacobi identities by construction and include the usual combinations in terms of master numerators as a special case. Our results provide an efficient algorithm, which is implemented in a {\sc Mathematica} package, for computing all tree amplitudes in theories including non-linear sigma model, special Galileon, Yang-Mills-scalar, Einstein-Yang-Mills and Dirac-Born-Infeld.}

\begin{document}
\maketitle
\addtocontents{toc}{\protect\setcounter{tocdepth}{1}}

\numberwithin{equation}{section}

		\tikzset{
		particles/.style={dashed, postaction={decorate},
			decoration={markings,mark=at position .5 with {\arrow[scale=1.5]{>}}
		}}
	}
	\tikzset{
		particle/.style={draw=black, postaction={decorate},
			decoration={markings,mark=at position .5 with {\arrow[scale=1.1]{>}}
		}}
	}
	\def  \layersep {.6cm}

\section{Introduction}
	
In 2008, Bern, Carrasco, Johansson (BCJ) proposed a remarkable duality between color and kinematics for scattering amplitudes in gauge theories, and double copy relations for corresponding gravity theories~\cite{Bern:2008qj, Bern:2010ue, Bern:2017yxu} (see~\cite{Bern:2019prr} for a review). The duality states that in a cubic-graph expansion of gauge-theory amplitudes (loop integrands), kinematic numerators can be arranged to satisfy the same relations as color factors, {\it i.e. Jacobi identities}; based on this  loop integrand for gravity amplitudes can be obtained by simply substituting color factors for another copy of such numerators. Although easy to prove at tree level~\cite{Bern:2010yg}, loop-level double-copy remains 
a conjecture, but it has led to great advances in the study of ultraviolet behavior of supergravity amplitudes
({\it c.f.} \cite{Bern:2012uf, Bern:2012cd, Bern:2013uka, Bern:2014sna, Bern:2017ucb, Bern:2018jmv}).
	
On the other hand, the Cachazo-He-Yuan (CHY) formulation \cite{Cachazo:2013hca, Cachazo:2013iea} has been one of the driving  forces in understanding and extending BCJ duality even at loop level. Based on scattering equations \cite{Cachazo:2013gna}, CHY formulas express tree amplitudes in various massless theories as integrals over moduli space ${\cal M}_{0,n}$ of certain worldsheet integrands, which can be derived from ambitwistor string theories \cite{Mason:2013sva} (see also \cite{Berkovits:2013xba, Adamo:2015hoa,Casali:2015vta}). This has not only led to new double-copy realizations and connections for theories \cite{Cachazo:2014xea,Cachazo:2014nsa}, but also extended BCJ double copy to one-loop level~\cite{He:2016mzd,He:2017spx, Edison:2020uzf,Edison:2020ehu} using loop-level CHY/ambitwistor strings \cite{Adamo:2013tsa,Geyer:2015bja,		Geyer:2015jch,Geyer:2016wjx,Geyer:2017ela,Geyer:2018xwu, Geyer:2019hnn}. Based on nodal Riemann spheres, loop-level ambitwistor-string formulas yield loop amplitudes in a new representation of Feynman integrals with propagators linear in loop momenta; alternatively they can be understood as forward limits of tree amplitudes with a pair of momenta in higher dimensions~\cite{He:2015yua, Cachazo:2015aol}. Everything we study at tree-level in this note can be uplifted to one loop with such forward limits.

It is an interesting question how to obtain closed-form expressions for such kinematic numerators in various theories, for example by taking field-theory limit of superstring amplitudes~\cite{Mafra:2011nw,Mafra:2011nv}. An alternative method, which has been studied extensively in the literature ({\it c.f.} \cite{Cachazo:2013iea,Bjerrum-Bohr:2016juj,Bjerrum-Bohr:2016axv,Du:2017kpo,Fu:2017uzt,Teng:2017tbo,Du:2017gnh,Edison:2020ehu,Huang:2017ydz,Cardona:2016gon}), is to extract them directly from CHY formulas, which gives a cubic-tree expansion with numerators automatically satisfying BCJ duality:
\be
	{\cal M}_{\rm tree}^{L \otimes R}=\sum_g \frac{{\bf n}^L_g {\bf n}^R_g}{\prod_I s_I}\,, \quad \textrm{with Jacobi-satisfying}~{\bf n}^L_g~\&~{\bf n}^R_g,
\ee
where the sum is over $(2n-5)!!$ cubic tree graphs labelled by $g$, and for each $g$, we have $(n-3)$ poles $s_I$, and ``numerators'' ${\bf n}^L_g$ and ${\bf n}^R_g$ which satisfy Jacobi-identities for any triplet of graphs~\cite{Bern:2008qj}. Such ${\bf n}^L_g$ and ${\bf n}^R_g$ can be read off from the two half-integrands of CHY formula (which indicates the theory), ${\cal I}_L$ and ${\cal I}_R$, respectively. As a trivial example, when one of the half-integrands is the ``color-dressed'' Parke-Taylor factor, (for particles in the adjoint of a gauge/flavor group) ,  ${\cal C}_n=\sum_{\pi \in S_{n-1}} {\rm PT}(\pi(1), \cdots, \pi(n)) {\rm Tr}(\pi)$ where ${\rm Tr}(\pi):={\rm Tr}(T^{a_{\pi(1)}} \cdots T^{a_{\pi(n)}})$ with the Parke-Taylor factor for a given ordering defined by 
	\be
	{\rm PT}(1,2,\cdots, n)=\frac 1 {\sigma_{1,2} \cdots \sigma_{n-1, n} \sigma_{n,1}}\,, \quad {\rm with}~\sigma_{i,j}:=\sigma_i -\sigma_j\,.
	\ee 
	Here ${\bf n}_g$ is the color factor~\footnote{We use a unified notation, ${\bf n}$ for both kinematic numerators and color ones, depending on what $L$ and $R$ means; for bi-adjoint $\phi^3$ theory, both are taken to be color/flavor factors.} for graph $g$, which is a linear combination of those for $(n{-}2)!$ half-ladder diagrams in DDM basis~\cite{DelDuca:1999rs}: ${\cal C}_n=\sum_{\pi\in S_{n{-}2}} c_{g(\pi)} {\rm PT}(1, \pi(2), \cdots,$ $ \pi(n{-}1), n)$. For half-integrands other than ${\cal C}_n$, it is non-trivial to extract kinematic numerators, which are required to be {\it local},  {\it i.e.} free of poles in Mandelstam variables. Systematic algorithms for	doing this based on scattering equations have been provided in {\it e.g}~\cite{He:2018pol,He:2019drm} (which also gives  double copy to string amplitudes). More recently a {\sc Mathematica} package is provided in \cite{Edison:2020ehu}, where a half-integrand can be expanded onto $(n{-}2)!$ Parke-Taylor factors in DDM basis, and the kinematic numerator ${\bf n}_g$ for any $g$ is given by the same linear combination of the $(n{-}2)!$ master numerators ${\bf n}_{g(\pi)}$ as for the color factors. A more invariant way for extracting  kinematic numerators is to directly take residues of half-integrands/string correlators in the moduli space ${\cal M}_{0,n}$~\cite{Mizera:2019blq} (see also \cite{Frost:2019fjn}).
	
In this  note we review and elaborate on an algorithm for efficiently computing all kinematic numerators ${\bf n}_g$ for any half integrand without the need of first computing master numerators. This is based on a natural expansion onto the so-called ``Cayley functions''~\cite{Gao:2017dek}, which are natural logarithmic functions on ${\cal M}_{0,n}$ in $1:1$ correspondence with $(n{-}1)^{n{-}3}$ {\it labelled trees} of $(n{-}1)$ points~\cite{Gao:2017dek}. By using the scattering-equation map~\cite{Arkani-Hamed:2017mur} to the space of Mandelstam variables, these functions give the so-called ``Cayley polytopes''~\cite{Gao:2017dek, Arkani-Hamed:2017mur} (see also \cite{He:2018pue,Feng:2020opo}), which have nice geometric interpretations and generalize the kinematic associahedra (mapped from Parke-Taylor factors). It turns out that any half-integrand with certain physical properties can be naturally expanded in these (vastly over-complete) worldsheet functions with relatively compact coefficients, which we call ``Cayley numerators''. The main result of this note is to give a general formula of ${\bf n}_g$ (for any cubic tree graph $g$) as a linear combination of Cayley numerators (for labelled trees), with $\pm 1$ coefficients. 

Without loss of generality, we can fix $\sigma_n \to\infty$, and for each labelled tree $\Gamma$ of $1,2,\cdots, n{-}1$, we define the {\it gauge-fixed} Cayley function $C_\Gamma$ to be
	\be
	C_\Gamma=\prod_{(i_a, j_a) \in E(\Gamma)}^{n{-}2} \frac 1{\sigma_{i_a, j_a}},
	\ee
	where the product is over all $n{-}2$ edges labelled by $(i_a, j_a)$ ($a=1,2,\cdots, n{-}2$)~\footnote{We have not specified {\it directions} of the edges, or orientation of $\Gamma$, thus $C_\Gamma$ is defined up to an overall sign, which will be discussed below. }, and the collection of all edges is denoted as $E(\Gamma)$. It is trivial to recover the dependence on $\sigma_n$ so we will also use $C_\Gamma$ to denote a Cayley function with $\sigma_n$ recovered. Note that $(n{-}1)!/2$ Parke-Taylor factors are special cases of Cayley functions, where $\Gamma$ are Hamilton graphs. It is well known~\cite{He:2018pol,He:2019drm} that on the support of scattering equations, any known half integrand can be written as a sum of Cayley functions, with coefficients (dubbed Cayley numerators) that turn out to be relatively compact functions of kinematics:
	\be\label{integrandexpansion}
	{\cal I}_n \simeq \sum_\Gamma {\bf n}_\Gamma (\epsilon, k) C_\Gamma\,.
	\ee

Note that such an expansion is of course not unique, and a special case is the expansion into $(n{-}2)!$ Parke-Taylor factors with coefficients being the  $(n{-}2)!$ BCJ numerators in DDM basis (even such an expansion is not unique since the $(n{-}2)!$ Parke-Taylor factors, while being linearly independent in general, satisfy linear relations implied by scattering equations). In this case, the Cayley numerators are non-zero only for $(n{-}2)!$ Hamilton graphs, $H(\pi)$ with $\pi \in S_{n{-}2}$, where $C_{H(\pi)}={\rm PT}(1, \pi(2), \cdots, \pi(n{-}1), n)$, and our result reduces to the well-known linear combinations of ${\bf n}_{H(\pi)}$ (in DDM basis) for all the ${\bf n}_g$'s~\footnote{These linear relations of Cayley functions which reduce any of them to a sum of Parke-Taylor factors, were first studied in~\cite{Stieberger:2013hza}.} However, we emphasize that our result is generic, {\it i.e.} it applies to any expansion of half-integrand into Cayley functions, and in the follow we find a natural expansion where all Cayley numerators turn out to be basic building blocks of kinematic numerators!

As will be reviewed in sec.~\ref{section3}, ${\bf n}_\Gamma$ become simple building blocks  since for all cases except that of ${\rm Pf}' \Psi_n$ they are {\it monomials} of kinematic variables (for ${\rm Pf}' \Psi_n$, a Cayley numerator is a sum of $n{-}1$ terms~\cite{Du:2017kpo}). This is in spirit very similar to the method of \cite{Lam:2018tgm}: instead of organizing the amplitude using ``denominators'' (propagators), the Cayley expansion suggests to organize it using these (monomial) ``numerators". We will use ${\rm det}' A_n$, ${\rm Pf}' \Psi_n$ and its dimension reduction {\it i.e. } ${\rm Pf} X_n {\rm Pf}' A_n$ as the main examples, which amounts to all kinematic numerators for tree amplitudes in NLSM, DBI (with photons and scalars), special Galileon (sGal), Yang-Mills-scalar (YMs), gravity (GR) and Einstein-Maxwell-scalar(EMs) theory. Explicitly, we will provide {\sc Mathematica} code for Cayley numerators and kinematic numerators from all these half-integrands and for those ``non-Abelian/squeezing'' generalization for generalized YMs, Einstein-Yang-Mills (EYM) {\it etc.}~\footnote{For YM, GR and generalizations such as EYM, BCJ master numerators are provided in {\it e.g.}~\cite{Du:2017gnh, He:2019drm, Edison:2020ehu}, and for NLSM~\cite{Du:2016tbc, Carrasco:2016ldy} but our method directly connects Cayley numerators to kinematic numerators.}. We remark that this way of computing tree amplitudes is rather efficient. For example, using our code on a laptop, it takes a second for computing the $8$-point amplitude in sGal, and one minute for $8$-point amplitudes in GR and BI (including those from dimension reduction)!

Before going to concrete theories, let us first present the general formula for writing kinematic numerator ${\bf n}_g$ for any cubic tree $g$ as a linear combinations of the ${\bf n}_\Gamma$'s, in sec~\ref{section2}.
	
\section{Kinematic numerators from Cayley numerators}\label{section2}
As shown in \cite{Gao:2017dek}, the CHY formula with any two Cayley functions gives a sum of cubic tree graphs with unit coefficients. In particular for $C_\Gamma^2$ we have
	\be\label{eq5}
	\int \dif \mu_n C_\Gamma^2= \sum_{g\in  T(\Gamma)} \prod_I \frac 1 {s_I}
	=\sum_{\substack{
			s_{I_1},s_{I_2},\cdots,s_{I_{n-3}} \in P(\Gamma)\\
			~\text{{\rm are}}~\text{{\rm compatible}~{\rm poles}}
	} }  \frac{1}{ s_{I_1}s_{I_2}\cdots s_{I_{n-3}}}\,.
	\ee
Here $T(\Gamma)$ denotes the collection of cubic trees (corresponding to vertices of the Cayley polytope~\cite{Gao:2017dek,Arkani-Hamed:2017mur,He:2018pue}), whose poles are kinematic invariants in the collection  $P(\Gamma)$  (corresponding to facets of the Cayley polytope) defined as 
	\ba\label{pc}
	P(\Gamma)=\Big\{s_{i_1,i_2,...,i_m}\Big|
	\raisebox{-.5cm}{
		\tikz[node distance =.5cm]{
			\node (a) {there is a connected sub-graph in the labelled tree of $\Gamma$   };
			\node [below of=a]{with nodes $\{i_1,i_2,...,i_m\}$ for $m=2,3,\cdots,n-2$};
	}}
	\Big\}\,.
	\ea
For example, for Hamilton tree of nodes $1,2,\cdots, n{-}1$, where $C_\Gamma={\rm PT}(1,2,\cdots,n)$, $P(\Gamma)$ is nothing but the collection of $n(n-3)/2$ planar variables (each corresponds to a subgraph of the $(n{-}1)$-point Hamilton tree), and $T(\Gamma)$ contains Calatan number of planar cubic trees; these are obviously familiar from the ABHY associahedra~\cite{Arkani-Hamed:2017mur}. Similarly we can define $P(\Gamma)$ and $T(\Gamma)$ for any labelled tree $\Gamma$. Moreover, for $\Gamma$ and $\Gamma'$, the CHY formula gives their ``intersection'':
	\be\label{mgamma}
	\int \dif \mu_n C_\Gamma C_{\Gamma'}=\pm \sum_{g\in  T(\Gamma)\cap T(\Gamma') } \prod_I \frac 1 {s_I}\equiv m(\Gamma|\Gamma')\,,
	\ee
where we sum over all cubic trees in the intersection of $T(\Gamma)$ and $T(\Gamma')$, and the overall sign is given in appendix \ref{signsign}. We have denoted the RHS as $m(\Gamma|\Gamma')$ as (vast) generalizations of double-partial amplitudes of bi-adjoint $\phi^3$, $m(\alpha|\beta)$ (where we have two Hamilton trees). We remark that the full amplitude is given by \be
M_{\rm tree}^{L\otimes R}=\sum_{\Gamma, \Gamma'} m(\Gamma|\Gamma') {\bf n}_\Gamma {\bf n}_{\Gamma'} \,.
\ee
In the rest of the note, we propose that instead of computing all $m(\Gamma|\Gamma')$, we simply compute ${\bf n}_g$ as linear combinations of ${\bf n}_\Gamma$: as $T(\Gamma)$ represents a map from any labelled tree to the collection of cubic trees, essentially what we are looking for is the inverse map.

\subsection{Cubic trees from labelled trees}
For this purpose, we introduce an abstract vector space where Cayley functions are vectors in it, and define their inner product via the CHY formula above. The map to cubic trees above can be viewed as an expansion
	\be\label{cayleymap}
	C_\Gamma \to  | \Gamma \>=  \sum_{g\in T(\Gamma)}    |g \>\<g|\Gamma\> 
	\ee
into  $(2n-5)!!$ auxiliary vectors labelled by cubic trees, $|g\>$, with inner product defined abstractly as
	\be\label{orthogonal}
	\<g|g'\>:= \delta_{g,g'} \prod_I \frac 1 {s_I}\,.
	\ee 
This means that they are orthogonal to each other and normalized to give a single cubic tree graph. It is easy to see that $\<g|\Gamma\>$ is nothing but a sign  since we must have 
	\ba\label{cayleysign}
	\int \dif \mu_n C_\Gamma C_{\Gamma'}= \sum_{g\in  T(\Gamma), g'\in  T({\Gamma'})  }  \<g|\Gamma\> \<g'|\Gamma'\>      \< g|g' \> =  \pm  \sum_{g\in  T(\Gamma)\cap T(\Gamma') }    \< g|g \>\,.
	\ea
Now we can consider any general half-integrand: given any expansion into Cayley functions \eqref{integrandexpansion}, using \eqref{cayleymap}, we have
\ba\label{newmap}
	|{\cal I}_n \>  = \sum_\Gamma {\bf n}_\Gamma  |\Gamma\> 
	=  \sum_\Gamma {\bf n}_\Gamma
\sum_{g\in T(\Gamma)}    |g \>\<g|\Gamma\> 
	= \sum_g  \left( \sum_{\Gamma \in F(g)} \<g|\Gamma\>   {\bf n}_\Gamma
  \right)
|g \>\,,
\ea 
where in the last equality, we have rearranged the summation: given $T(\Gamma)$ as the collection of cubic trees produced by any labelled tree $\Gamma$ by \eqref{eq5}, we define $F(g)$ as the collection of labelled trees that can produce $g$ through \eqref{eq5}. As a main result of this section, we give the construction of $F(g)$ in sec~\ref{secFg}. The linear combination of ${\bm n}_\Gamma$ in the parenthesis is nothing but a BCJ-satisfying numerator of a cubic tree $g$
\ba \label{ng1}
{\bm n}_g= \sum_{\Gamma \in F(g)} \<g|\Gamma\>   {\bf n}_\Gamma \,,
\ea 
since the CHY integral of two half-integrands simply becomes
	\ba\label{InInp}
	\int \dif \mu_n   {\cal I}_n^L {\cal I}_n^R=\<{\cal I}_n^L | {\cal I}_n^R \>=  \sum_{g , g'}  {\bm n}_g^L {\bm n}_{g'}^R        \< g|g' \> =  \sum_{g }  {\bm n}_g^L {\bm n}_{g}^R  \prod_I \frac 1 {s_I}   \,,  
	\ea
where in the last equality we have used the orthogonality relations \eqref{orthogonal}.

Before we give the explicit construction of $F(g)$, let's first see how to fix the signs $\<g|\Gamma\>$ in the summation of \eqref{ng1}. It turns out  to be a   convenient way to introduce a planar ordering $\rho=(\rho(1),\rho(2),\cdots,\rho(n-1),n)$ consistent with the cubic tree $g$, where $\rho(1,2,\cdots, n{-}1)$ can also be viewed as a Hamilton tree; After pulling out an overall sign $\<g | \rho\>$, which is not important since it
 squares to $1$ and drops in \eqref{InInp} by making use of the orthogonality relations  \eqref{orthogonal}, all the relative signs can be nicely determined as:
\ba \label{ng}
\boxed{
{\bm n}_g= \<g|\rho\>  \sum_{\Gamma \in F(g)}    {\rm sgn}^\rho_\Gamma \, {\bf n}_\Gamma \,.
}
\ea
More details for this formula are given in appendix \ref{signsign}. Here $ {\rm sgn}_{\Gamma}^{\rho} $ is given by the product 
	\be\label{sign}
	{\rm sgn}_{\Gamma }^{\rho}= \prod_{(i_a, j_a) \in E(\Gamma)}^{n{-}2}   {\rm sgn}_{i_a,j_a}^{\rho}
	\ee
of signs associated with edges of the labelled tree $\Gamma$, defined with respect to the ordering $\rho$:
	\ba
	\operatorname{sgn}_{i j}^{\rho}=\left\{\begin{array}{l}+1: i \text { is on the left of } j \text { in } \rho(1,2, \ldots, n-1) \\ -1: i \text { is on the right of } j \text { in } \rho(1,2, \ldots, n-1)\end{array}\right.
	\,.
	\ea
For example we have
	$	\operatorname{sgn}^{123}_{23}=1,
	\operatorname{sgn}^{123}_{31}=-1,
	\operatorname{sgn}^{132}_{12}=1
	$.
	
Note that for each cubic tree $g$, we can choose its own ordering $\rho$ in \eqref{ng}: for a color ordered amplitude, there is a universal ordering for all planar cubic trees but this is not true for general cases. For example, for $n=4$ with the $3$ cubic trees in $s,t,u$-channel respectively, one cannot find a universal $\rho$ since they do not share a common planar ordering.
	
Before proceeding to the construction of $F(g)$ in general, let's spell out a $n=4$ example in full detail. Consider the half-integrand ${\rm det}' A_n$ (needed for NLSM, DBI and special Galileon) whose definition can be found in sec \ref{section3}. Now for $n=4$ with $\s_4\to\infty$, it is trivial to expand it into the $3$ labelled trees with remarkably simple numerators:
\ba\label{present}
{\rm det}'{A_4} \simeq \frac{s_{12}s_{23} }{\s_{12}\s_{23}}+\frac{s_{13}s_{23} }{\s_{13}\s_{32}}+\frac{s_{12}s_{13} }{\s_{12}\s_{13}}\,,
\ea 
where the tree Cayley numerators are given by monomials of Mandelstam variables
\ba \label{eq20}
{\bf n}_{1\gets2\gets3}= s_{12}s_{23}\,,\quad 
{\bf n}_{1\gets3\gets2}= s_{13}s_{23}\,,\quad 
{\bf n}_{3\to1\gets2}= s_{12}s_{13}\,,\quad
\ea 
associated with edges of the labelled trees; we draw the $3$ trees (denoted as $1\gets2\gets3$ {\it etc.} above) explicitly as (with $1$ as the root for orientation):
\ba\label{eq19}
\raisebox{-10pt}{
\tikz[scale=.5]{	\filldraw (0,0) circle(2pt)  node[below=1.5pt]{$1$};
\filldraw (1,0) circle(2pt)  node[below=1.5pt]{$2$};
\filldraw (2,0) circle(2pt)  node[below=1.5pt]{$3$};
	\draw[particle] (1,0)--(0,0);
	\draw[particle] (2,0)--(1,0);
}}\,,\quad 
\raisebox{-10pt}{
\tikz[scale=.5]{	\filldraw (0,0) circle(2pt)  node[below=1.5pt]{$1$};
\filldraw (1,0) circle(2pt)  node[below=1.5pt]{$3$};
\filldraw (2,0) circle(2pt)  node[below=1.5pt]{$2$};
	\draw[particle] (1,0)--(0,0);
	\draw[particle] (2,0)--(1,0);
	}
}\,,\quad 
\raisebox{-10pt}{
\tikz[scale=.5]{	\filldraw (0,0) circle(2pt)  node[below=1.5pt]{$3$};
\filldraw (1,0) circle(2pt)  node[below=1.5pt]{$1$};
\filldraw (2,0) circle(2pt)  node[below=1.5pt]{$2$};
	\draw[particle] (0,0)--(1,0);
	\draw[particle] (2,0)--(1,0);
}
}
\,.
\ea 
We proceed to find the kinematic numerators in terms of these Cayley numerators. By $s\!\!:=s_{12},\, t\!\!:=s_{23},\, u\!\!:=s_{13}$ (recall $s_{i j}:=k_i\cdot k_j$), we use $g_s,g_t,g_u$ to denote the $3$ cubic trees. What we want to find is the expansion 
\ba
|{\rm det}{}'A_4\>=
n_s |g_s\>  + n_t |g_t\>  + n_u |g_u\>  
\,,
\ea
where ${ n}_s,{ n}_t,{ n}_u$ are BCJ-satisfying kinematic numerators. It is trivial for $n=4$ to obtain cubic trees from labelled trees:
\ba 
T(
\raisebox{-10pt}{
\tikz[scale=.5]{	\filldraw (0,0) circle(2pt)  node[below=1.5pt]{$1$};
\filldraw (1,0) circle(2pt)  node[below=1.5pt]{$2$};
\filldraw (2,0) circle(2pt)  node[below=1.5pt]{$3$};
	\draw[particle] (1,0)--(0,0);
	\draw[particle] (2,0)--(1,0);
}
}
)=\{g_s,g_t\}\,,\quad T(
\raisebox{-10pt}{
\tikz[scale=.5]{	\filldraw (0,0) circle(2pt)  node[below=1.5pt]{$1$};
\filldraw (1,0) circle(2pt)  node[below=1.5pt]{$3$};
\filldraw (2,0) circle(2pt)  node[below=1.5pt]{$2$};
	\draw[particle] (1,0)--(0,0);
	\draw[particle] (2,0)--(1,0);
}
}
)=\{g_t,g_u\}\,,\quad T(
\raisebox{-10pt}{
\tikz[scale=.5]{	\filldraw (0,0) circle(2pt)  node[below=1.5pt]{$3$};
\filldraw (1,0) circle(2pt)  node[below=1.5pt]{$1$};
\filldraw (2,0) circle(2pt)  node[below=1.5pt]{$2$};
	\draw[particle] (0,0)--(1,0);
	\draw[particle] (2,0)--(1,0);
}
}
)=\{g_s,g_u\}\,.
\ea thus we can immediately find all the $F(g)$, which are collections of  unoriented labelled trees: 
\ba \label{eqfgs}
& F(g_s)=\{\!\! \raisebox{-10pt}{
\tikz[scale=.45]{	\filldraw (0,0) circle(2pt)  node[below=1.5pt]{$1$};
\filldraw (1,0) circle(2pt)  node[below=1.5pt]{$2$};
\filldraw (2,0) circle(2pt)  node[below=1.5pt]{$3$};
	\draw (1,0)--(0,0);
	\draw (2,0)--(1,0);
}
}\!\!
,\!\!
\raisebox{-10pt}{
\tikz[scale=.45]{	\filldraw (0,0) circle(2pt)  node[below=1.5pt]{$2$};
\filldraw (1,0) circle(2pt)  node[below=1.5pt]{$1$};
\filldraw (2,0) circle(2pt)  node[below=1.5pt]{$3$};
	\draw (0,0)--(1,0);
	\draw (2,0)--(1,0);
}
}
\!\!\}\,,\; 
F(g_t) =\{ \!\!
\raisebox{-10pt}{
\tikz[scale=.45]{	\filldraw (0,0) circle(2pt)  node[below=1.5pt]{$1$};
\filldraw (1,0) circle(2pt)  node[below=1.5pt]{$2$};
\filldraw (2,0) circle(2pt)  node[below=1.5pt]{$3$};
	\draw (1,0)--(0,0);
	\draw (2,0)--(1,0);
}
}\!\!
,
\!\!\raisebox{-10pt}{
\tikz[scale=.45]{	\filldraw (0,0) circle(2pt)  node[below=1.5pt]{$1$};
\filldraw (1,0) circle(2pt)  node[below=1.5pt]{$3$};
\filldraw (2,0) circle(2pt)  node[below=1.5pt]{$2$};
	\draw (1,0)--(0,0);
	\draw (2,0)--(1,0);
}
}
 \!\!\}\,,\;
F(g_u) =\{\!\! \raisebox{-10pt}{
\tikz[scale=.45]{	\filldraw (0,0) circle(2pt)  node[below=1.5pt]{$1$};
\filldraw (1,0) circle(2pt)  node[below=1.5pt]{$3$};
\filldraw (2,0) circle(2pt)  node[below=1.5pt]{$2$};
	\draw (0,0)--(1,0);
	\draw (1,0)--(2,0);
}
}\!\!
,\!\!
\raisebox{-10pt}{
\tikz[scale=.45]{	\filldraw (0,0) circle(2pt)  node[below=1.5pt]{$2$};
\filldraw (1,0) circle(2pt)  node[below=1.5pt]{$1$};
\filldraw (2,0) circle(2pt)  node[below=1.5pt]{$3$};
	\draw (1,0)--(0,0);
	\draw (1,0)--(2,0);
}
}
\!\!\}\,.
\ea 
Now we are ready to obtain ${ n}_s, { n}_t, { n}_u$ according to \eqref{ng}.  We can choose $\rho=123$ for $g_s$ since $(1234)$ is a planar ordering of the $s$-channel cubic tree $g_s$.  So we have
\ba 
{n}_{s}= \<g|123\>  \sum_{\Gamma \in F(g_s)}    {\rm sgn}^{123}_\Gamma \, { n}_\Gamma= \<g_s|123\>  \left(  {\rm sgn}^{123}_ {1\gets2\gets3} \, { n}_{1\gets2\gets3} 
+
 {\rm sgn}^{123}_ {3\to1\gets2} \, { n}_{3\to1\gets2} 
\right)\,,
\ea 
with
$
 {\rm sgn}^{123}_ {1\gets2\gets3}=  {\rm sgn}^{123}_ {12}  {\rm sgn}^{123}_ {23}=1 
 $ and $
  {\rm sgn}^{123}_ {1\gets2,1\gets3}=  {\rm sgn}^{123}_ {12}  {\rm sgn}^{123}_ {13}=1
  $. 
This leads to 
\ba \label{eq25}
n_s= \<g_s|123\>  s_{12}( s_{23}+ s_{13} )\,,\quad 
n_t= \<g_t|123\>  s_{23}( s_{12}- s_{13} )\,,\quad 
n_u= \<g_u|132\>  s_{13}( s_{23}+ s_{12} )\,,\quad 
\ea 
where the other two numerators are obtained similarly. 

Four-point tree amplitude of sGal is given by the CHY formula with two copies of ${\rm det}{}'A_4$ \cite{Cachazo:2014xea}:
\ba \label{sgal4}
M_4^{\rm sGal}=\<{\rm det}{}'A_4 |{\rm det}{}'A_4\>= \frac{n_s^2}{s}+ \frac{n_t^2}{t}+ \frac{n_u^2}{u}=s^3+t(s-u)^2+u^3=-s t u\,.
\ea 
Similarly by pairing it with a Parke-Taylor factor {\it e.g.} with ordering $(1234)$:
\ba 
|{\rm PT}(1,2,3,4)\>= \<g_s|123\> |g_s\> + \<g_t|123\> |g_t\>\,,
\ea
we obtain the correct color-ordered amplitude in NLSM
\ba \label{nlsm4}
M^{\rm NLSM}[1234]=\< {\rm PT}(1,2,3,4)| {\rm det}{}'A_4\> =\frac{n_s}{s}+ \frac{n_t}{t}=-s +(s-u)=-u\,.
\ea

 \subsection{General construction of $F(g)$ \label{secFg}}
 
Now we present our construction for $F(g)$, which is the most important ingredient to obtain kinematic numerator $n_g$ in \eqref{ng}.
Note that \eqref{pc} requires that there is a sub-graph in every labelled tree $\Gamma\in F(g)$ with nodes $\{i_1,i_2,...,i_m\}$ for every 
pole $s_{i_1,i_2,\cdots,i_m}$ of $g$; this means we can construct each element of $F(g)$ by adding $n{-}2$ edges recursively: we start from the $n{-}1$ nodes with no edges at all, and in each step we produce a collection of {\it forests} (with node $1,2,\cdots, n-1$) which has exactly one more edge than before, until the last step when all of them are connected to give the collection of labelled trees. 

To state the rule in a universal manner, in addition to the $n{-}3$ poles of $g$, which we can write as $s_I$ for $I \subset \{1,2,\dots, n{-}1\}$, we also introduce an additional pole $s_{1,2,\cdots, n{-}1}$. We then add $n{-}2$ edges recursively according to these $n{-}2$ $s_I$ with $|I|=2,\cdots, n{-}1$:
\begin{itemize}
    \item Start from $n-1$ nodes $1,2,\cdots,n-1$. For each two-particle pole $s_{i,j}$, we have a forest with one edge connecting $i$ and $j$. 
    \item For any multi-particle pole $s_I$, there must be either two poles $s_{I_1}$, $s_{I_2}$ with $I_1\sqcup I_2=I$, or a pole with exactly one less point, $s_{I/\{i\}}$; in the first case, we have a forest with one more edge connecting $i_1\in I_1, i_2\in I_2$, and in the second case, we have a forest with one more edge connecting $i$ and $j\in I/\{i\}$. 
    
    \item The same applies to $s_{1,2,\cdots, n{-}1}$: after $n-3$ steps we have a collection of forests where each of them has $n{-}3$ edges, thus exactly two disconnected subgraphs, and $F(g)$ is given by adding the last edge connecting them in all possible ways. 
\end{itemize}
The orientations of edges of the labelled trees only affect the overall sign of the Cayley function which drops via the map \eqref{eq5}. We use an unoriented labelled tree to denote the collection of oriented ones with all possible directions on each edge.
 
For example, $F(g_s)$ from \eqref{eqfgs} can be obtained following this procedure:
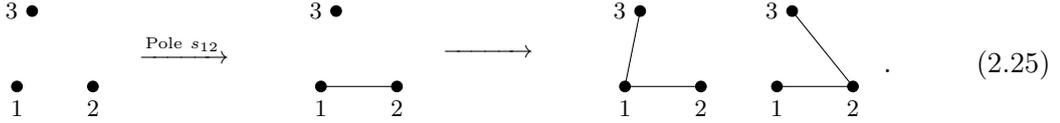
\ba 
\raisebox{-20pt}{
\begin{tikzpicture}[every node/.style={font=\footnotesize,},vertex/.style={inner sep=0,minimum size=3pt,circle,fill},wavy/.style={decorate,decoration={coil,aspect=0, segment length=2mm, amplitude=0.5mm}},dir/.style={decoration={markings, mark=at position \halfwayb with {\arrow[scale=0.5]{Latex}}},postaction={decorate}},scale=1]
\begin{scope}[xshift=0cm,yshift=0cm]
\filldraw (0,0) circle(2pt)  node[below=1.5pt]{$1$}; 
\filldraw (1,0) circle(2pt)  node[below=1.5pt]{$2$}; 
\filldraw (.2,1) circle(2pt)  node[left=1.5pt]{$3$}; 
\node at (2.2,0.5) {$\xrightarrow{{\rm Pole} ~ s_{12} }$};
\end{scope}
\begin{scope}[xshift=4cm,yshift=0cm]
\filldraw (0,0) circle(2pt)  node[below=1.5pt]{$1$}; 
\filldraw (1,0) circle(2pt)  node[below=1.5pt]{$2$}; 
\filldraw (.2,1) circle(2pt)  node[left=1.5pt]{$3$}; 
\draw (0,0)--(1,0);
\node at (2.2,0.5) {$\xrightarrow{ ~~~~~~~~~ }$}; 
\end{scope}
\begin{scope}[xshift=8cm,yshift=0cm]
\filldraw (0,0) circle(2pt)  node[below=1.5pt]{$1$}; 
\filldraw (1,0) circle(2pt)  node[below=1.5pt]{$2$}; 
\filldraw (.2,1) circle(2pt)  node[left=1.5pt]{$3$}; 
\draw (0,0)--(1,0);
\draw (0,0)--(.2,1);
\end{scope}
\begin{scope}[xshift=10cm,yshift=0cm]
\filldraw (0,0) circle(2pt)  node[below=1.5pt]{$1$}; 
\filldraw (1,0) circle(2pt)  node[below=1.5pt]{$2$}; 
\filldraw (.2,1) circle(2pt)  node[left=1.5pt]{$3$}; 
\draw (0,0)--(1,0);
\draw (1,0)--(.2,1);
\end{scope}
\end{tikzpicture}
}
\,.
\ea 
Let us also present several examples for $n=5,6$, and in general such $F(g)$ can be easily produced by our {\sc Mathematica} notebook. 

For the $n=5$ cubic tree with poles $s_{2 3},s_{1 2 3}$, our procedure gives $6$ labelled trees:
  \ba
  \raisebox{0pt}
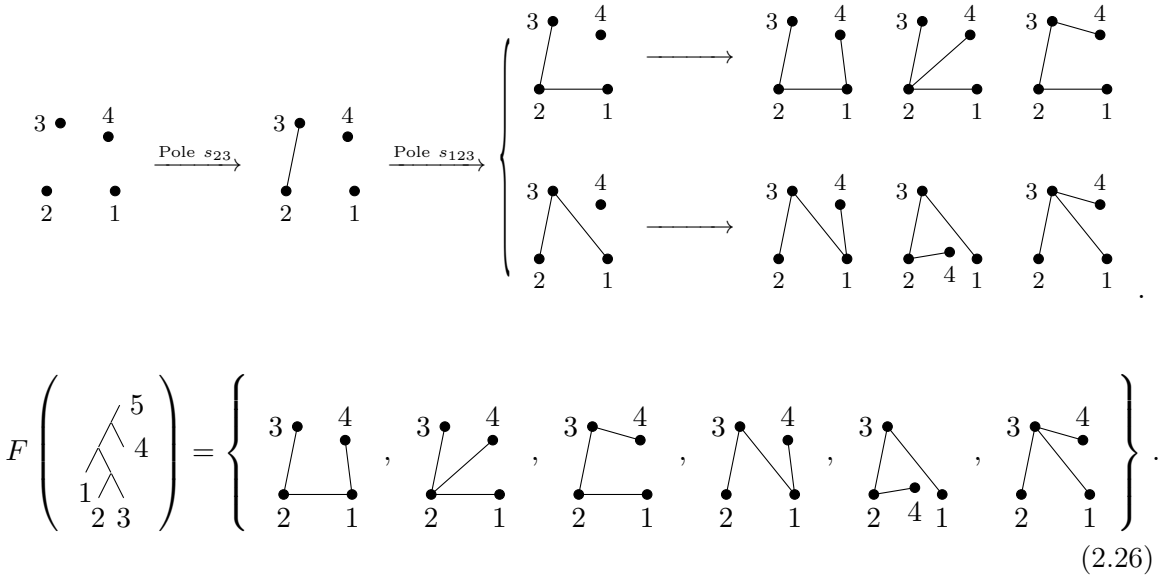
{
\begin{tikzpicture}[every node/.style={font=\footnotesize,},vertex/.style={inner sep=0,minimum size=3pt,circle,fill},wavy/.style={decorate,decoration={coil,aspect=0, segment length=2mm, amplitude=0.5mm}},dir/.style={decoration={markings, mark=at position \halfwayb with {\arrow[scale=0.5]{Latex}}},postaction={decorate}},scale=1]
\begin{scope}[scale=.9,xshift=-3.5cm,yshift=0cm]
\filldraw (0,0) circle(2pt)  node[below=1.5pt]{$2$}; 
\filldraw (1,0) circle(2pt)  node[below=1.5pt]{$1$}; 
\filldraw (.2,1) circle(2pt)  node[left=1.5pt]{$3$}; 
\filldraw (.9,.8) circle(2pt)  node[above=1.5pt]{$4$}; 
\node at (2.2,0.5) {$\xrightarrow{{\rm Pole}~s_{2 3} }$}; 
\end{scope}
\begin{scope}[scale=.9,xshift=0cm,yshift=0cm]
\filldraw (0,0) circle(2pt)  node[below=1.5pt]{$2$}; 
\filldraw (1,0) circle(2pt)  node[below=1.5pt]{$1$}; 
\filldraw (.2,1) circle(2pt)  node[left=1.5pt]{$3$}; 
\filldraw (.9,.8) circle(2pt)  node[above=1.5pt]{$4$}; 
\draw (0,0)--(.2,1);
\node at (2.2,0.5) {$\xrightarrow{{\rm Pole}~s_{123} }$}; 
\node at (3.2,0.5) {$\stretchto[2000]{\{}{90pt}$};
\end{scope}
\begin{scope}[scale=.9,xshift=3.7cm,yshift=1.5cm]
\filldraw (0,0) circle(2pt)  node[below=1.5pt]{$2$}; 
\filldraw (1,0) circle(2pt)  node[below=1.5pt]{$1$}; 
\filldraw (.2,1) circle(2pt)  node[left=1.5pt]{$3$}; 
\filldraw (.9,.8) circle(2pt)  node[above=1.5pt]{$4$}; 
\draw (0,0)--(.2,1);
\draw (0,0)--(1,0);
\node at (2.2,0.5) {$\xrightarrow{ ~~~~~~~~~ }$}; 
\end{scope}
\begin{scope}[scale=.9,xshift=3.7cm,yshift=-1cm]
\filldraw (0,0) circle(2pt)  node[below=1.5pt]{$2$}; 
\filldraw (1,0) circle(2pt)  node[below=1.5pt]{$1$}; 
\filldraw (.2,1) circle(2pt)  node[left=1.5pt]{$3$}; 
\filldraw (.9,.8) circle(2pt)  node[above=1.5pt]{$4$}; 
\draw (0,0)--(.2,1);
\draw (.2,1)--(1,0);
\node at (2.2,0.5) {$\xrightarrow{ ~~~~~~~~~ }$}; 
\end{scope}
\begin{scope}[scale=.9,xshift=7.2cm,yshift=1.5cm]
\filldraw (0,0) circle(2pt)  node[below=1.5pt]{$2$}; 
\filldraw (1,0) circle(2pt)  node[below=1.5pt]{$1$}; 
\filldraw (.2,1) circle(2pt)  node[left=1.5pt]{$3$}; 
\filldraw (.9,.8) circle(2pt)  node[above=1.5pt]{$4$}; 
\draw (0,0)--(.2,1);
\draw (0,0)--(1,0);
\draw (1,0)--(.9,.8);
\end{scope}
\begin{scope}[scale=.9,xshift=9.1cm,yshift=1.5cm]
\filldraw (0,0) circle(2pt)  node[below=1.5pt]{$2$}; 
\filldraw (1,0) circle(2pt)  node[below=1.5pt]{$1$}; 
\filldraw (.2,1) circle(2pt)  node[left=1.5pt]{$3$}; 
\filldraw (.9,.8) circle(2pt)  node[above=1.5pt]{$4$}; 
\draw (0,0)--(.2,1);
\draw (0,0)--(1,0);
\draw (0,0)--(.9,.8);
\end{scope}
\begin{scope}[scale=.9,xshift=11.0cm,yshift=1.5cm]
\filldraw (0,0) circle(2pt)  node[below=1.5pt]{$2$}; 
\filldraw (1,0) circle(2pt)  node[below=1.5pt]{$1$}; 
\filldraw (.2,1) circle(2pt)  node[left=1.5pt]{$3$}; 
\filldraw (.9,.8) circle(2pt)  node[above=1.5pt]{$4$}; 
\draw (0,0)--(.2,1);
\draw (0,0)--(1,0);
\draw (.2,1)--(.9,.8);
\end{scope}
\begin{scope}[scale=.9,xshift=7.2cm,yshift=-1cm]
\filldraw (0,0) circle(2pt)  node[below=1.5pt]{$2$}; 
\filldraw (1,0) circle(2pt)  node[below=1.5pt]{$1$}; 
\filldraw (.2,1) circle(2pt)  node[left=1.5pt]{$3$}; 
\filldraw (.9,.8) circle(2pt)  node[above=1.5pt]{$4$}; 
\draw (0,0)--(.2,1);
\draw (.2,1)--(1,0);
\draw (1,0)--(.9,.8);
\end{scope}
\begin{scope}[scale=.9,xshift=9.1cm,yshift=-1cm]
\filldraw (0,0) circle(2pt)  node[below=1.5pt]{$2$}; 
\filldraw (1,0) circle(2pt)  node[below=1.5pt]{$1$}; 
\filldraw (.2,1) circle(2pt)  node[left=1.5pt]{$3$}; 
\filldraw (.6,.1) circle(2pt)  node[below=1.5pt]{$4$}; 
\draw (0,0)--(.2,1);
\draw (.2,1)--(1,0);
\draw (0,0)-- (.6,.1) ;
\end{scope}
\begin{scope}[scale=.9,xshift=11.0cm,yshift=-1cm]
\filldraw (0,0) circle(2pt)  node[below=1.5pt]{$2$}; 
\filldraw (1,0) circle(2pt)  node[below=1.5pt]{$1$}; 
\filldraw (.2,1) circle(2pt)  node[left=1.5pt]{$3$}; 
\filldraw (.9,.8) circle(2pt)  node[above=1.5pt]{$4$}; 
\draw (0,0)--(.2,1);
\draw (.2,1)--(1,0);
\draw (.2,1)--(.9,.8);
\end{scope}
\end{tikzpicture}
}
\,.\nonumber
\ea  
   \ba 
   \label{5ptex}
 F\left( \raisebox{-29pt}{
  \begin{tikzpicture}[scale=.9,shorten >=0pt,draw=black,scale=.25,
        node distance = .2cm,
        neuron3/.style = {circle, minimum size=.1pt, inner sep=0pt,  fill=black } ]
     \node[neuron3] {}
     child {node[neuron3] {} 
                 child {node[neuron3] {}node[below=0pt]{$1$}}
        child {node[neuron3] {}
         child {node[neuron3] {}node[below=0pt]{$2$}}
            child {node[neuron3] {}node[below=0pt]{$3$}}
        }
        }
        child {node[neuron3] {}
        node[right=0pt]{$4$}
        }
           ;
     \draw (0,0)--(.5,1) node[right=0pt]{$5$};
    \end{tikzpicture}} 
    \right)=
    \left\{
   \raisebox{-29pt}{
  \begin{tikzpicture}[scale=.9]
\filldraw (0,0) circle(2pt)  node[below=1.5pt]{$2$}; 
\filldraw (1,0) circle(2pt)  node[below=1.5pt]{$1$}; 
\filldraw (.2,1) circle(2pt)  node[left=1.5pt]{$3$}; 
\filldraw (.9,.8) circle(2pt)  node[above=1.5pt]{$4$}; 
\draw (0,0)--(.2,1);
\draw (0,0)--(1,0);
\draw  (1,0)--(.9,.8);
\end{tikzpicture}
}\,,
   \raisebox{-29pt}{
  \begin{tikzpicture}[scale=.9]
\filldraw (0,0) circle(2pt)  node[below=1.5pt]{$2$}; 
\filldraw (1,0) circle(2pt)  node[below=1.5pt]{$1$}; 
\filldraw (.2,1) circle(2pt)  node[left=1.5pt]{$3$}; 
\filldraw (.9,.8) circle(2pt)  node[above=1.5pt]{$4$}; 
\draw (0,0)--(.2,1);
\draw (0,0)--(1,0);
\draw (0,0)--(.9,.8);
\end{tikzpicture}
}\,,
   \raisebox{-29pt}{
  \begin{tikzpicture}[scale=.9]
\filldraw (0,0) circle(2pt)  node[below=1.5pt]{$2$}; 
\filldraw (1,0) circle(2pt)  node[below=1.5pt]{$1$}; 
\filldraw (.2,1) circle(2pt)  node[left=1.5pt]{$3$}; 
\filldraw (.9,.8) circle(2pt)  node[above=1.5pt]{$4$}; 
\draw (0,0)--(.2,1);
\draw (0,0)--(1,0);
\draw  (.2,1)--(.9,.8);
\end{tikzpicture}
}\,,
   \raisebox{-29pt}{
  \begin{tikzpicture}[scale=.9]
\filldraw (0,0) circle(2pt)  node[below=1.5pt]{$2$}; 
\filldraw (1,0) circle(2pt)  node[below=1.5pt]{$1$}; 
\filldraw (.2,1) circle(2pt)  node[left=1.5pt]{$3$}; 
\filldraw (.9,.8) circle(2pt)  node[above=1.5pt]{$4$}; 
\draw  (0,0)--(.2,1);
\draw (.2,1)--(1,0);
\draw (1,0)--(.9,.8);
\end{tikzpicture}
}\,,
   \raisebox{-29pt}{
  \begin{tikzpicture}[scale=.9]
\filldraw (0,0) circle(2pt)  node[below=1.5pt]{$2$}; 
\filldraw (1,0) circle(2pt)  node[below=1.5pt]{$1$}; 
\filldraw (.2,1) circle(2pt)  node[left=1.5pt]{$3$}; 
\filldraw (.6,.1) circle(2pt)  node[below=1.5pt]{$4$}; 
\draw (0,0)--(.2,1);
\draw  (.2,1)--(1,0);
\draw (0,0)-- (.6,.1) ;
\end{tikzpicture}
}\,,
   \raisebox{-29pt}{
  \begin{tikzpicture}[scale=.9]
\filldraw (0,0) circle(2pt)  node[below=1.5pt]{$2$}; 
\filldraw (1,0) circle(2pt)  node[below=1.5pt]{$1$}; 
\filldraw (.2,1) circle(2pt)  node[left=1.5pt]{$3$}; 
\filldraw (.9,.8) circle(2pt)  node[above=1.5pt]{$4$}; 
\draw  (0,0)--(.2,1);
\draw (.2,1)--(1,0);
\draw  (.2,1)--(.9,.8);
\end{tikzpicture}
}
    \right\}\,.
\ea 
For the $n=6$ cubic tree with poles $s_{12},s_{34},s_{1234}$, the procedure gives
	  
	  \ba \label{eq34}
\begin{tikzpicture}[every node/.style={font=\footnotesize,},vertex/.style={inner sep=0,minimum size=3pt,circle,fill},wavy/.style={decorate,decoration={coil,aspect=0, segment length=2mm, amplitude=0.5mm}},dir/.style={decoration={markings, mark=at position \halfwayb with {\arrow[scale=0.5]{Latex}}},postaction={decorate}},scale=1]
\begin{scope}[xshift=0cm,yshift=0cm]
\filldraw (0,0) circle(2pt)  node[below=1.5pt]{$2$}; 
\filldraw (1,0) circle(2pt)  node[below=1.5pt]{$1$}; 
\filldraw (.2,1) circle(2pt)  node[left=1.5pt]{$3$}; 
\filldraw (.9,.8) circle(2pt)  node[above=1.5pt]{$4$}; 
\filldraw (.5,.5) circle(2pt)  node[left=1.5pt]{$5$}; 
\draw (0,0)--(1,0);
\draw (.2,1)--(.9,.8);
\node at (2.2,0.5) {$\xrightarrow{{\rm Pole}~s_{1234} }$}; 
\node at (3.2,0.5) {$\stretchto[2000]{\{}{150pt}$};
\end{scope}
\begin{scope}[xshift=3.7cm,yshift=2cm]
\filldraw (0,0) circle(2pt)  node[below=1.5pt]{$2$}; 
\filldraw (1,0) circle(2pt)  node[below=1.5pt]{$1$}; 
\filldraw (.2,1) circle(2pt)  node[left=1.5pt]{$3$}; 
\filldraw (.9,.8) circle(2pt)  node[above=1.5pt]{$4$}; 
\filldraw (.5,.5) circle(2pt)  node[left=1.5pt]{$5$}; 
\draw (0,0)--(1,0);
\draw (.2,1)--(.9,.8);
\draw (1,0)--(.9,.8);
\node at (2.2,0.5) {$\xrightarrow{ ~~~~~~~~~} $}; 
\end{scope}
\begin{scope}[xshift=3.7cm,yshift=0cm]
\filldraw (0,0) circle(2pt)  node[below=1.5pt]{$2$}; 
\filldraw (1,0) circle(2pt)  node[below=1.5pt]{$1$}; 
\filldraw (.2,1) circle(2pt)  node[left=1.5pt]{$3$}; 
\filldraw (.9,.8) circle(2pt)  node[above=1.5pt]{$4$}; 
\filldraw (.3,.5) circle(2pt)  node[left=1.5pt]{$5$}; 
\draw (0,0)--(1,0);
\draw (.2,1)--(.9,.8);
\draw (1,0)-- (.2,1);
\node at (2.2,0.5) {$\xrightarrow{ ~~~~~~~~~} $}; 
\end{scope}
\begin{scope}[xshift=7.2cm,yshift=2cm]
\filldraw (0,0) circle(2pt)  node[below=1.5pt]{$2$}; 
\filldraw (1,0) circle(2pt)  node[below=1.5pt]{$1$}; 
\filldraw (.2,1) circle(2pt)  node[left=1.5pt]{$3$}; 
\filldraw (.9,.8) circle(2pt)  node[above=1.5pt]{$4$}; 
\filldraw (.5,.5) circle(2pt)  node[left=1.5pt]{$5$}; 
\draw (0,0)--(1,0);
\draw (.2,1)--(.9,.8);
\draw (1,0)--(.9,.8);
\draw (1,0)--(.5,.5);
\end{scope}
\begin{scope}[xshift=9.1cm,yshift=2cm]
\filldraw (0,0) circle(2pt)  node[below=1.5pt]{$2$}; 
\filldraw (1,0) circle(2pt)  node[below=1.5pt]{$1$}; 
\filldraw (.2,1) circle(2pt)  node[left=1.5pt]{$3$}; 
\filldraw (.9,.8) circle(2pt)  node[above=1.5pt]{$4$}; 
\filldraw (.5,.5) circle(2pt)  node[left=1.5pt]{$5$}; 
\draw (0,0)--(1,0);
\draw (.2,1)--(.9,.8);
\draw (1,0)--(.9,.8);
\draw (0,0)--(.5,.5);
\end{scope}
\begin{scope}[xshift=11.0cm,yshift=2cm]
\filldraw (0,0) circle(2pt)  node[below=1.5pt]{$2$}; 
\filldraw (1,0) circle(2pt)  node[below=1.5pt]{$1$}; 
\filldraw (.2,1) circle(2pt)  node[left=1.5pt]{$3$}; 
\filldraw (.9,.8) circle(2pt)  node[above=1.5pt]{$4$}; 
\filldraw (.5,.5) circle(2pt)  node[left=1.5pt]{$5$}; 
\draw (0,0)--(1,0);
\draw (.2,1)--(.9,.8);
\draw (1,0)--(.9,.8);
\draw (.2,1)--(.5,.5);
\end{scope}
\begin{scope}[xshift=12.9cm,yshift=2cm]
\filldraw (0,0) circle(2pt)  node[below=1.5pt]{$2$}; 
\filldraw (1,0) circle(2pt)  node[below=1.5pt]{$1$}; 
\filldraw (.2,1) circle(2pt)  node[left=1.5pt]{$3$}; 
\filldraw (.9,.8) circle(2pt)  node[above=1.5pt]{$4$}; 
\filldraw (.5,.5) circle(2pt)  node[left=1.5pt]{$5$}; 
\draw (0,0)--(1,0);
\draw (.2,1)--(.9,.8);
\draw (1,0)--(.9,.8);
\draw (.9,.8)--(.5,.5);
\end{scope}
\begin{scope}[xshift=7.2cm,yshift=0cm]
\filldraw (0,0) circle(2pt)  node[below=1.5pt]{$2$}; 
\filldraw (1,0) circle(2pt)  node[below=1.5pt]{$1$}; 
\filldraw (.2,1) circle(2pt)  node[left=1.5pt]{$3$}; 
\filldraw (.9,.8) circle(2pt)  node[above=1.5pt]{$4$}; 
\filldraw (.3,.5) circle(2pt)  node[left=1.5pt]{$5$}; 
\draw (0,0)--(1,0);
\draw (.2,1)--(.9,.8);
\draw (1,0)-- (.2,1);
\draw (1,0)-- (.3,.5);
\end{scope}
\begin{scope}[xshift=9.1cm,yshift=0cm]
\filldraw (0,0) circle(2pt)  node[below=1.5pt]{$2$}; 
\filldraw (1,0) circle(2pt)  node[below=1.5pt]{$1$}; 
\filldraw (.2,1) circle(2pt)  node[left=1.5pt]{$3$}; 
\filldraw (.9,.8) circle(2pt)  node[above=1.5pt]{$4$}; 
\filldraw (.3,.5) circle(2pt)  node[left=1.5pt]{$5$}; 
\draw (0,0)--(1,0);
\draw (.2,1)--(.9,.8);
\draw (1,0)-- (.2,1);
\draw (0,0)-- (.3,.5);
\end{scope}
\begin{scope}[xshift=11.0cm,yshift=0cm]
\filldraw (0,0) circle(2pt)  node[below=1.5pt]{$2$}; 
\filldraw (1,0) circle(2pt)  node[below=1.5pt]{$1$}; 
\filldraw (.2,1) circle(2pt)  node[left=1.5pt]{$3$}; 
\filldraw (.9,.8) circle(2pt)  node[above=1.5pt]{$4$}; 
\filldraw (.3,.5) circle(2pt)  node[left=1.5pt]{$5$}; 
\draw (0,0)--(1,0);
\draw (.2,1)--(.9,.8);
\draw (1,0)-- (.2,1);
\draw (.2,1)-- (.3,.5);
\end{scope}
\begin{scope}[xshift=12.9cm,yshift=0cm]
\filldraw (0,0) circle(2pt)  node[below=1.5pt]{$2$}; 
\filldraw (1,0) circle(2pt)  node[below=1.5pt]{$1$}; 
\filldraw (.2,1) circle(2pt)  node[left=1.5pt]{$3$}; 
\filldraw (.9,.8) circle(2pt)  node[above=1.5pt]{$4$}; 
\filldraw (.8,.5) circle(2pt)  node[right=1.5pt]{$5$}; 
\draw (0,0)--(1,0);
\draw (.2,1)--(.9,.8);
\draw (1,0)-- (.2,1);
\draw (.9,.8)-- (.8,.5);
\end{scope}
\begin{scope}[xshift=6.2cm,yshift=-2.2cm]
\node at (2.2,0.5) {$\,\,\Bigg(1\xleftrightarrow{ ~~~~~~~} 2\Bigg) $}; 
\end{scope}
\end{tikzpicture}
\ea  
where we abbreviated $8$ trees obtained from the first two lines with $1$ and $2$ exchanged, and in total we have 16 (unoriented) labelled trees in $F\left(
\raisebox{-29pt}{
  \begin{tikzpicture}[shorten >=0pt,draw=black,scale=.25,
        node distance = .2cm,
        neuron3/.style = {circle, minimum size=.1pt, inner sep=0pt,  fill=black } ]
     \node[neuron3] {}
     child {node[neuron3] {} 
                 child {node[neuron3] {}
             child {node[neuron3] {}  node[below=0pt]{$1$} }
                child {node[neuron3] {} node[below=0pt]{$2$}  }     
                           }
                child[white] {node[neuron3] {}}               child {node[neuron3] {}
         child {node[neuron3] {}node[below=0pt]{$3$}}
            child {node[neuron3] {}node[below=0pt]{$4$}}
        }
        }
        child {node[neuron3] {}
        node[right=0pt]{$5$}
        }
           ;
     \draw (0,0)--(.5,1) node[right=0pt]{$6$};
    \end{tikzpicture}} 
\right)
$.	

We provide a function 
$
\texttt {F}[poles\_]
$ 
in the auxiliary {\sc Mathematica} 
package that automatically generates the labelled tree set $F(g)$. Here the cubic tree $g$ is given by its $n{-}3$ poles. For example, 
$ F\left(
 	\raisebox{-29pt}{
  \begin{tikzpicture}[shorten >=0pt,draw=black,scale=.25,
        node distance = .2cm,
        neuron3/.style = {circle, minimum size=.1pt, inner sep=0pt,  fill=black } ]
     \node[neuron3] {}
node[neuron3] {} 
                 child {node[neuron3] {}
             child {node[neuron3] {}  node[below=0pt]{$1$} }
                child {node[neuron3] {} node[below=0pt]{$2$}  }     
                           }
                child[white] {node[neuron3] {}}               child {node[neuron3] {}
         child {node[neuron3] {}node[below=0pt]{$3$}}
            child {node[neuron3] {}node[below=0pt]{$4$}}
        }
           ;
     \draw (0,0)--(.5,1) node[right=0pt]{$5$};
    \end{tikzpicture}} 
    \right)$ is given by 
\ba  \label{5ptex2}
\texttt {F}[s_{1,2}s_{3,4}]=&\Big\{
\big\{\{2,1\},\{3,2\},\{4,3\}\big\},
\big\{\{2,1\},\{3,1\},\{4,3\}\big\},
\nl&
\,\,\big\{\{2,1\},\{3,4\},\{4,1\}\big\},
\big\{\{2,1\},\{4,2\},\{3,4\}\big\}
\Big\},
\nl
\texttt {F}[s_{1,2}s_{3,4},\textsl {Graph} \to \textsl {True}]=&  \left\{
   \raisebox{-29pt}{
  \begin{tikzpicture}
\filldraw (0,0) circle(2pt)  node[below=1.5pt]{$2$}; 
\filldraw (1,0) circle(2pt)  node[below=1.5pt]{$1$}; 
\filldraw (.2,1) circle(2pt)  node[left=1.5pt]{$3$}; 
\filldraw (.9,.8) circle(2pt)  node[above=1.5pt]{$4$}; 
\draw (0,0)--(.2,1);
\draw (0,0)--(1,0);
\draw  (.2,1)--(.9,.8);
\end{tikzpicture}
}\;,
\raisebox{-29pt}{
\begin{tikzpicture}
\filldraw (0,0) circle(2pt)  node[below=1.5pt]{$2$}; 
\filldraw (1,0) circle(2pt)  node[below=1.5pt]{$1$}; 
\filldraw (.2,1) circle(2pt)  node[left=1.5pt]{$3$}; 
\filldraw (.9,.8) circle(2pt)  node[above=1.5pt]{$4$}; 
\draw (1,0)--(.2,1);
\draw (0,0)--(1,0);
\draw  (.2,1)--(.9,.8);
\end{tikzpicture}
}\;,
\raisebox{-29pt}{
\begin{tikzpicture}
\filldraw (0,0) circle(2pt)  node[below=1.5pt]{$2$}; 
\filldraw (1,0) circle(2pt)  node[below=1.5pt]{$1$}; 
\filldraw (.2,1) circle(2pt)  node[left=1.5pt]{$3$}; 
\filldraw (.9,.8) circle(2pt)  node[above=1.5pt]{$4$}; 
\draw (1,0)--(.9,.8);
\draw (0,0)--(1,0);
\draw  (.2,1)--(.9,.8);
\end{tikzpicture}
}\;,
\raisebox{-29pt}{
\begin{tikzpicture}
\filldraw (0,0) circle(2pt)  node[below=1.5pt]{$2$}; 
\filldraw (1,0) circle(2pt)  node[below=1.5pt]{$1$}; 
\filldraw (.2,1) circle(2pt)  node[left=1.5pt]{$3$}; 
\filldraw (.9,.8) circle(2pt)  node[above=1.5pt]{$4$}; 
\draw (0,0)--(.9,.8);
\draw (0,0)--(1,0);
\draw  (.2,1)--(.9,.8);
\end{tikzpicture}
}\;
\right\}\,.
\ea 
Here any unoriented labelled tree is indicated by the collection of edges, and the option ${\textsl {Graph}}\to \textsl {True}$ allows us to see them explicitly. We also provide 
\ba 
&
\texttt {AllCubicTrees}[n\_],\quad \texttt {PlanarCubicTrees}[ordering\_] ,\quad 
\texttt {AllLabelledTrees}[n\_]\,,
\ea 
to generate all cubic trees, all planar cubic trees and all labelled trees respectively. We remind the reader that one can simply learn about any {\sc Mathematica} function appeared in this note by using $?$ in the code. For example, 
\ba\label{inquiring} 
\texttt{?F}= &\text{  Generates a labelled tree set $F(g)$.
}\\&\quad
e.g.\quad  \texttt{F}[s_{1,2}s_{3,4}], \quad \texttt{F}[s_{1,2}s_{3,4}s_{5,6},\textsl {Graph} \to \textsl {True}],\quad   \texttt{F}[s[2,3]s[1,2,3]s[1,2,3,4]]
\nl
&\quad{\rm option}: \textsl {Graph} \to \textsl {True},\, \textsl {Root} \to \textsl {1}
\nl
&\quad{\text {see also}}: 
\texttt{AllLabelledTrees},\quad \texttt{nCayleyNLSM},\quad \texttt{nCayleyYM}
\nonumber
\ea
where we provide examples for the usage and several related functions.

	
\section{Expansion of half-integrands and explicit kinematic numerators in various theories}\label{section3}

We have presented the abstract formula \eqref{ng}, for kinematic numerators in terms of Cayley numerators for any half-integrand, where the key ingredient is the general construction of $F(g)$.  In this section, we move to specific half integrands and find explicit kinematic numerators by using the expansion into Cayley functions. Such expansions have been obtained for essentially all CHY half-integrands known so far, and for simplicity we focus on ${\rm det}' A_n$, ${\rm Pf}' \Psi_n$ and its dimension reduction. 
We  provide an auxiliary {\sc Mathematica} notebook which gives general results for tree amplitudes in all theories with CHY formulas. 
	
\subsection{${\rm det'}A_n$: NLSM and sGal}
	
Let us start with the simplest case. For NLSM and sGal, all we need (in addition to Parke-Taylor factor) is the reduced determinant 
	\ba \label{deta}
	{\rm det}{}'{ A}_n={(-1)^{i+n}} {\rm det}|{ A}|_{i}^{i}\,,
	\ea 
	where ${ A}_n$ is a $(n-1)\times (n-1)$ symmetric matrix with ${ A}_{a,a}=0 $ and
	$
	{ A}_{a,b}= \frac{s_{a,b}}{\s_{a,b}}$ for $ 1\leq a\neq b\leq n-1$.
	$\s_n$ is absent since we send it to infinity and delete the $n$-th column and row. 
The matrix has one additional null vector on the support of scattering equations. It is convenient to rewrite the diagonal elements using them
\ba
A_{a,a}\simeq-\sum_{\substack{b=1\\b\neq a} }^{n-1}\frac{s_{a,b}}{\s_{a,b}}\,,
\ea 
which makes it manifest that entries along any column/row add up to zero. The reduced determinant is defined by deleting any column and row, and by the matrix-tree theorem ({\it c.f.} \cite{Feng:2012sy}) it is given by the sum over all labelled trees with nodes $1,2,\cdots, n{-}1$, where the summand is simply the product of $A_{i,j}$ for all the $n{-}2$ edges $(i,j)$. Therefore, we have a natural expansion of $\det' A_n$ into Cayley functions:
\ba \label{eq35}
{\rm det}{}'{A_n}  \simeq \sum_{\Gamma} {\bf n}_\Gamma C_{\Gamma} \,,
\ea
where the orientation of any tree is determined by (all arrows flowing into) a root which we choose to be node $1$ (the root depends on which row and column we delete in \eqref{deta}; see \eqref{eq19} for an $n=4$ example). Each Cayley numerator ${\bf n}_\Gamma$ is simply the product of $n{-}2$ Mandelstam variables for the edges of the labelled tree $\Gamma$:
	\be\label{ngamma}
	\boxed{
	n_{\Gamma }= \prod_{(i_a, j_a) \in E(\Gamma)}^{n{-}2} s_{i_a,j_a}\,.
	}
	\ee 
This illustrates nicely how Cayley numerators are simple building blocks: not only each ${\bf n}_\Gamma$ is a monomial of Mandelstam variables, but also the expansion is  manifestly symmetric in $2,3,\cdots, n{-}1$. By \eqref{ng}, we trivially obtain kinematic numerators for any cubic trees for NLSM or sGal amplitudes to all $n$ (see \eqref{eq20} for the $n=4$ example)! Note that for odd $n$, ${\rm det}{}'A_n$ vanishes trivially but \eqref{eq35} and \eqref{ngamma} still hold: it is just that the full amplitude vanishes by momentum conservation.

We remark that this result of course implies the well-known expressions of BCJ numerators found in~\cite{Du:2016tbc, Carrasco:2016ldy}. Take the half-ladder diagram with ordering $2,3,\cdots, n{-}1$ as an example. It is trivial to see that for this cubic tree, we generate $(n{-}2)!$ labelled trees in $F(g)$: the first edge must be $2\dedge1$ (for $s_{12}$ pole), the second one $3\dedge1$ or $3\dedge2$ (for $s_{123}$), then $4\dedge1$, $4\dedge2$ or $4\dedge3$ (for $s_{1234}$), {\it etc.}, until the last one connecting $n{-}1$ to any of $1,2,\cdots, n{-}2$  (for $s_{12\cdots n{-}1})$; for these $(n{-}2)!$ $\Gamma$ with root $1$, by choosing $\rho=12\cdots n{-}1$, the sign of every edge is always $-1$, thus we have
\be
{n}_g=(-1)^n\, s_{12} (s_{13}+s_{23}) \cdots (s_{1,n-1}+s_{2,n-1}+ \cdots + s_{n-2 ,n{-}1})\,.
\ee
Now our formula also generates a similar but more involved expression for numerators of other cubic trees that are not in the DDM basis. 

Let us give some explicit examples for $n=5,6$. Since there is a manifest symmetry for $2,3,\cdots, n{-}1$ in our construction, we only need to consider numerators for a small subset of $(2n-5)!!$ cubic trees since others can be obtained by relabelling; in practice this is extremely useful for computing the amplitude in NLSM and sGal. For example, for $n=5$, it suffices to consider numerators of the following $4$ (out of $15$) cubic trees:

\vspace{-.5cm}
\ba \label{4cubic}
	\raisebox{-25pt}{
  \begin{tikzpicture}[shorten >=0pt,draw=black,scale=.25,
        node distance = .2cm,
        neuron3/.style = {circle, minimum size=.1pt, inner sep=0pt,  fill=black } ]
     \node[neuron3] {}
node[neuron3] {} 
                 child {node[neuron3] {}
             child {node[neuron3] {}  node[below=0pt]{$1$} }
                child {node[neuron3] {} node[below=0pt]{$2$}  }     
                           }
                child[white] {node[neuron3] {}}               child {node[neuron3] {}
         child {node[neuron3] {}node[below=0pt]{$3$}}
            child {node[neuron3] {}node[below=0pt]{$4$}}
        }
           ;
     \draw (0,0)--(.5,1) node[right=0pt]{$5$};
    \end{tikzpicture}} 
    \quad
 \raisebox{-25pt}{
  \begin{tikzpicture}[shorten >=0pt,draw=black,scale=.25,
        node distance = .2cm,
        neuron3/.style = {circle, minimum size=.1pt, inner sep=0pt,  fill=black } ]
     \node[neuron3] {}
     child {node[neuron3] {} 
                 child {node[neuron3] {}node[below=0pt]{$1$}}
        child {node[neuron3] {}
         child {node[neuron3] {}node[below=0pt]{$2$}}
            child {node[neuron3] {}node[below=0pt]{$3$}}
        }
        }
        child {node[neuron3] {}
        node[right=0pt]{$4$}
        }
           ;
     \draw (0,0)--(.5,1) node[right=0pt]{$5$};
    \end{tikzpicture}} 
    \quad 
    \raisebox{-25pt}{
  \begin{tikzpicture}[shorten >=0pt,draw=black,scale=.25,
        node distance = .2cm,
        neuron3/.style = {circle, minimum size=.1pt, inner sep=0pt,  fill=black } ]
     \node[neuron3] {}
     child {node[neuron3] {} 
                 child {node[neuron3] {}
                          child {node[neuron3] {}node[below=0pt]{$1$}}
            child {node[neuron3] {}node[below=0pt]{$2$}}
}
        child {node[neuron3] {}             
    node[right=0pt]{$3$}
        }
        }
        child {node[neuron3] {}
        node[right=0pt]{$4$}
        }
           ;
     \draw (0,0)--(.5,1) node[right=0pt]{$5$};
    \end{tikzpicture}} 
   \quad 
       \raisebox{-25pt}{
  \begin{tikzpicture}[shorten >=0pt,draw=black,scale=.25,
        node distance = .2cm,
        neuron3/.style = {circle, minimum size=.1pt, inner sep=0pt,  fill=black } ]
     \node[neuron3] {}
     child {node[neuron3] {} 
     node[below=0pt]{$1$}
        }
        child {node[neuron3] {}
                 child {node[neuron3] {}node[below=0pt]{$2$}}
        child {node[neuron3] {}
         child {node[neuron3] {}node[below=0pt]{$3$}}
            child {node[neuron3] {}node[below=0pt]{$4$}}
        }
        }
           ;
     \draw (0,0)--(.5,1) node[right=0pt]{$5$};
    \end{tikzpicture}} 
    \,.
\ea 
According to \eqref{5ptex2}, using \eqref{ng} we have
\ba 
n\left(\!\!\!	\raisebox{-23pt}{
  \begin{tikzpicture}[scale=.8, shorten >=0pt,draw=black,scale=.25,
        node distance = .2cm,
        neuron3/.style = {circle, minimum size=.1pt, inner sep=0pt,  fill=black } ]
     \node[neuron3] {}
node[neuron3] {} 
                 child {node[neuron3] {}
             child {node[neuron3] {}  node[below=0pt]{$1$} }
                child {node[neuron3] {} node[below=0pt]{$2$}  }     
                           }
                child[white] {node[neuron3] {}}               child {node[neuron3] {}
         child {node[neuron3] {}node[below=0pt]{$3$}}
            child {node[neuron3] {}node[below=0pt]{$4$}}
        }
           ;
     \draw (0,0)--(.5,1) node[right=0pt]{$5$};
    \end{tikzpicture}} 
   \!\! \right)= 
   \pm\left[
-{\bf n}\left(\!\!\!
   \raisebox{-23pt}{
  \begin{tikzpicture}[scale=.8]
\filldraw (0,0) circle(2pt)  node[below=1.5pt]{$2$}; 
\filldraw (1,0) circle(2pt)  node[below=1.5pt]{$1$}; 
\filldraw (.2,1) circle(2pt)  node[left=1.5pt]{$3$}; 
\filldraw (.9,.8) circle(2pt)  node[above=1.5pt]{$4$}; 
\draw[particle] (.2,1)--(0,0);
\draw[particle] (0,0)--(1,0);
\draw[particle]  (.9,.8)--(.2,1);
\end{tikzpicture}
}
\!\!\right)-{\bf n}\left(\!\!\!
\raisebox{-23pt}{
\begin{tikzpicture}[scale=.8]
\filldraw (0,0) circle(2pt)  node[below=1.5pt]{$2$}; 
\filldraw (1,0) circle(2pt)  node[below=1.5pt]{$1$}; 
\filldraw (.2,1) circle(2pt)  node[left=1.5pt]{$3$}; 
\filldraw (.9,.8) circle(2pt)  node[above=1.5pt]{$4$}; 
\draw[particle] (.2,1)--(1,0);
\draw[particle] (0,0)--(1,0);
\draw[particle]  (.9,.8)--(.2,1);
\end{tikzpicture}
}
\!\!\right)+{\bf n}\left(\!\!\!
\raisebox{-23pt}{
\begin{tikzpicture}[scale=.8]
\filldraw (0,0) circle(2pt)  node[below=1.5pt]{$2$}; 
\filldraw (1,0) circle(2pt)  node[below=1.5pt]{$1$}; 
\filldraw (.2,1) circle(2pt)  node[left=1.5pt]{$3$}; 
\filldraw (.9,.8) circle(2pt)  node[above=1.5pt]{$4$}; 
\draw[particle] (.9,.8)--(1,0);
\draw[particle] (0,0)--(1,0);
\draw[particle]  (.2,1)--(.9,.8);
\end{tikzpicture}
}
\!\!\right)+{\bf n}\left(\!\!\!
\raisebox{-23pt}{
\begin{tikzpicture}[scale=.8]
\filldraw (0,0) circle(2pt)  node[below=1.5pt]{$2$}; 
\filldraw (1,0) circle(2pt)  node[below=1.5pt]{$1$}; 
\filldraw (.2,1) circle(2pt)  node[left=1.5pt]{$3$}; 
\filldraw (.9,.8) circle(2pt)  node[above=1.5pt]{$4$}; 
\draw[particle] (.9,.8)--(0,0);
\draw[particle] (0,0)--(1,0);
\draw[particle]  (.2,1)--(.9,.8);
\end{tikzpicture}
}
\!\!\right)
\right]\,,
\ea 
where we have written $n_g, {\bf n}_\Gamma$ as $n(g), {\bf n}(\Gamma)$ for readability, and use {\it e.g.} $n(s_{12}s_{34})$ to denote the kinematic numerator of the corresponding cubic tree. Besides, we simply write the sign  $\<s_{12}s_{34}| 1234\>$ as $\pm$ since it drops in the end. The Cayley numerator can be read off directly according to \eqref{ngamma}, and we obtain the kinematic numerator for this cubic tree
\ba 
n(s_{12}s_{34}) =\pm s_{12}s_{34}(-s_{23}-s_{13}+s_{14}+s_{24})\,.
 \ea 
Similarly we obtain numerators for the other $3$ cubic trees:
 \ba 
n(s_{23}s_{123}) =& \pm s_{23}s_{123}(s_{12}-s_{13})\,,
\qquad\qquad
n(s_{12}s_{123}) = \pm s_{12}s_{123}(s_{13}+s_{23})
\,,
\nl
n(s_{34}s_{234}) =& \pm s_{34}[s_{23}(s_{13}-s_{12}-s_{14})+s_{24}(s_{12}+s_{13}-s_{14})]
\,.
 \ea 
and with relabelling in $2,3,4$, they give all kinematic numerators for $n=5$ amplitudes in sGal and NLSM (which vanish by momentum conservation). 

We remark that \eqref{ngamma} and \eqref{ng} give explicit formulas for NLSM and sGal amplitudes to any multiplicity. Practically the computation is very efficient since we only need those distinct cubic trees under relabelling of $2,3,\cdots, n{-}1$; out of $(2n{-}5)!!$ trees, we only need $2, 4, 9, 20, 46, 106, 249$ such trees for $n=4,5,\cdots, 10$; once we write down the sum over these cubic trees, the full amplitude is simply given by adding permutations in $2,3,\cdots, n{-}1$.  Using our code on a laptop, it takes about one second for computing $8$-point sGal amplitude, and only a few minutes for $10$-point sGal.
 
More amplitudes and kinematic numerators in sGal or NLSM can be generated in the auxiliary {\sc Mathematica} file through the functions
\ba 
&{\texttt {AmpsGal}}[ n\_],\qquad\qquad
\texttt {nCubicsGal}[poles\_]
,
\nl
& \texttt {AmpNLSM}[ ordering\_]  \,, \quad 
\texttt {nCubicNLSM}[poles\_,ordering\_] ,
\nonumber
\ea 
respectively. For example, one can use  $\texttt {AmpsGal}[ 4]$ and  $\texttt {AmpNLSM}[ \{1,2,3,4\}]$ to generate \eqref{sgal4} and \eqref{nlsm4} respectively.  $\texttt {AmpsGal}[n, \textsl {Symmetric} \to \textsl {True}]$ organizes the amplitude in a way where the permutation invariance w.r.t. $2,3,\cdots,n{-}1$ becomes manifest. It is much more efficient when $n$ is big.
$\texttt {nCubicNLSM}[s_{1,2}s_{3,4}$ $s_{1,2,3,4},$ $\{1,2,3,4,5,6\}]$ generates a numerator of partial NLSM amplitude  with ordering $(123456)$,
\ba \label{nlsm6}
\texttt {nCubicNLSM}[s_{1,2}s_{3,4}s_{1,2,3,4},\{1,2,3,4,5,6\}]
=
s_{12}s_{34}s_{1234}(-s_{23}-s_{13}+s_{14}+s_{24}) 
\,,
 \ea 
where in fact the code has directly used \eqref{eq34}. We also provide
\ba 
&
\texttt {nCayleyNLSM}[pairs\_\_] ,\qquad \texttt {nCubicNLSM}[poles\_]
\nonumber
\ea 
for Cayley numerators or corresponding cubic-tree numerators of ${\rm det}{}'A_n$. Due to \eqref{ngamma}, the function  $\texttt {nCayleyNLSM}[pairs\_\_]$ becomes trivial, {\it e.g.} 
\ba 
\texttt {nCayleyNLSM}[\{2,1\},\{3,2\},\{4,2\},\{5,3\}]= s_{12}s_{23}s_{24}s_{35}\,,
\ea 
and $\texttt {nCubicNLSM}[s_{1,2}s_{3,4}s_{1,2,3,4}]$ almost produces the same result as \eqref{nlsm6} except that an additional sign $\left\<
s_{12}s_{34}s_{1234}   | 12345 \right\>$ is added.



\subsection{${\rm Pf}' \Psi_n$ and dimension reduction: YM, GR, DBI {\it etc.}}
	
Next we move to the reduced Pfaffian, ${\rm Pf}' \Psi_n$, which gives YM (with Parke-Taylor) and GR, as well as Born-Infeld (BI) (with ${\rm det}' A_n$). The expansion into Cayley functions is well known~\cite{Du:2017kpo}, and the resulting Cayley numerators are significantly more complicated than that for ${\rm det}' A_n$ above. Thus we will only briefly review the algorithm for getting ${ n}_\Gamma$'s, and later we will study a special case more explicitly by taking the dimension reduction~\footnote{The details of ${\rm Pf}' \Psi_n$,  and its dimension reduction ${\rm Pf}' \Pi_n$ can be found in~\cite{Cachazo:2014xea}.}.

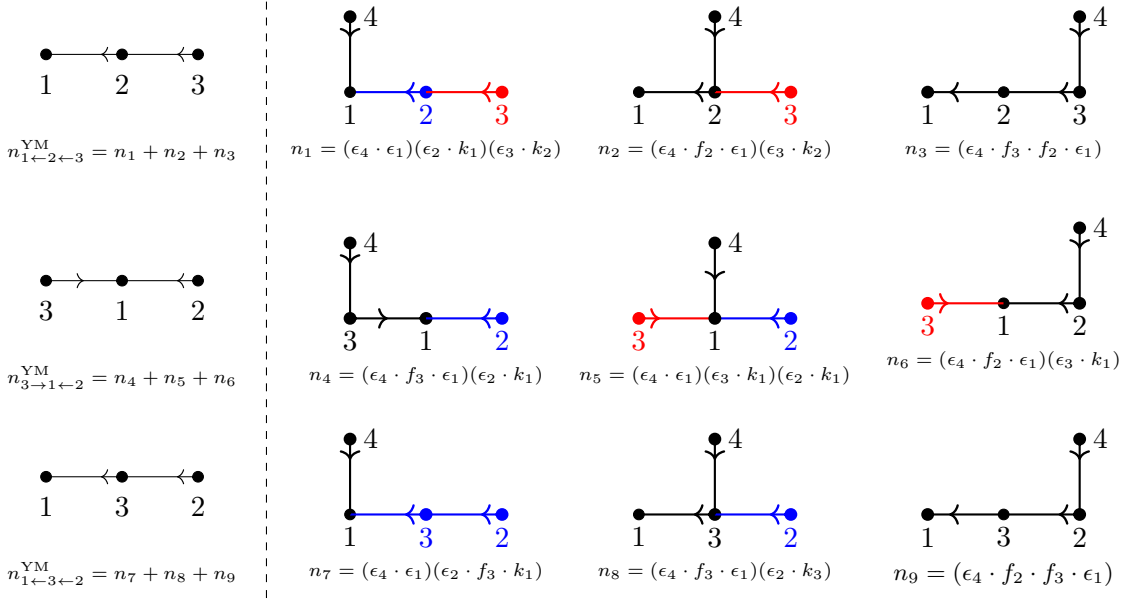
\begin{figure}[!h]
\centering
\begin{tikzpicture}[scale=1,shorten >=0pt,draw=black,
node distance = \layersep,
neuron/.style = {circle, minimum size=3pt, inner sep=0pt, fill=black } ]
\draw[dashed] (-2.1,1.2)--(-2.1,-6.8);
\begin{scope}[xshift=-4cm]
\coordinate (1) at (-1,0.5);
\coordinate (2) at (0,0.5);
\coordinate (3) at (1,0.5);
\filldraw (1) circle(2pt) node[below=4pt]{$1$};
\filldraw[particle] (2) circle(2pt) node[below=4pt]{$2$}--(1);
\filldraw[particle] (3) circle(2pt) node[below=4pt]{$3$} -- (2);
\node at (0,0) [below=0.5cm,font=\fontsize{7}{6}\selectfont]{${n}^{\rm YM}_{1\gets2\gets3}=n_1+n_2+n_3$};
\end{scope}
\begin{scope}
\coordinate (1) at (-1,0);
\coordinate (2) at (0,0);
\coordinate (3) at (1,0);
\coordinate (4) at (-1,1);
\filldraw [particle,color=blue,thick ] (2) circle(2pt) node[below=1pt]{$2$}--(1) ;
\filldraw [particle,color=red,thick ] (3) circle(2pt) node[below=1pt]{$3$}--(2);
\filldraw (1) circle(2pt) node[below=1pt]{$1$};
\filldraw [particle,thick] (4) circle(2pt) node[right=1pt]{$4$}--(1) ;
\node at (0,0) [below=0.5cm,font=\fontsize{7}{6}\selectfont]{$n_1=(\epsilon_4\cdot\epsilon_1)(\epsilon_2\cdot k_1)(\epsilon_3\cdot k_2)$};
\end{scope}
\begin{scope}[xshift=3.8cm]
\coordinate (1) at (-1,0);
\coordinate (2) at (0,0);
\coordinate (3) at (1,0);
\coordinate (4) at (0,1);
\filldraw (1) circle(2pt) node[below=1pt]{$1$};
\filldraw [particle,thick] (2) circle(2pt) node[below=1pt]{$2$}--(1);
\filldraw [particle,color=red,thick ] (3) circle(2pt) node[below=1pt]{$3$}--(2);
\filldraw [particle,thick] (4) circle(2pt) node[right=1pt]{$4$}--(2);
\node at (0,0) [below=0.5cm,font=\fontsize{7}{6}\selectfont]{$n_2=(\epsilon_4\cdot f_2 \cdot\epsilon_1)(\epsilon_3\cdot k_2)$};
\end{scope}
\begin{scope}[xshift=7.6cm]
\coordinate (1) at (-1,0);
\coordinate (2) at (0,0);
\coordinate (3) at (1,0);
\coordinate (4) at (1,1);
\filldraw (2) circle(2pt) node[below=1pt]{$2$};
\filldraw [particle,thick] (2)--(1) circle(2pt) node[below=1pt]{$1$};
\filldraw [particle,thick] (3) circle(2pt) node[below=1pt]{$3$}--(2);
\filldraw [particle,thick] (4) circle(2pt) node[right=1pt]{$4$}--(3);
\node at (0,0) [below=0.5cm,font=\fontsize{7}{6}\selectfont]{$n_3=(\epsilon_4\cdot f_3\cdot f_2\cdot\epsilon_1)$};
\end{scope}
\begin{scope}[xshift=-4cm,yshift=-3cm]
\coordinate (3) at (-1,0.5);
\coordinate (1) at (0,0.5);
\coordinate (2) at (1,0.5);
\filldraw[particle] (3) circle(2pt) node[below=4pt]{$3$}--(1) circle(2pt) node[below=4pt]{$1$};
\filldraw[particle] (2) circle(2pt) node[below=4pt]{$2$}--(1);
\node at (0,0) [below=0.5cm,font=\fontsize{7}{6}\selectfont]{${n}^{\rm YM}_{3\to1\gets2}=n_4+n_5+n_6$};
\end{scope}
\begin{scope}[yshift=-3cm]
\coordinate (3) at (-1,0);
\coordinate (1) at (0,0);
\coordinate (2) at (1,0);
\coordinate (4) at (-1,1);
\filldraw [particle,thick] (3) circle(2pt) node[below=1pt]{$3$}--(1) circle(2pt) node[below=1pt]{$1$};
\filldraw [particle,color=blue,thick ] (2) circle(2pt) node[below=1pt]{$2$}--(1);
\filldraw [particle,thick] (4) circle(2pt) node[right=1pt]{$4$}--(3);
\node at (0,0) [below=0.5cm,font=\fontsize{7}{6}\selectfont]{$n_4=(\epsilon_4\cdot f_3\cdot\epsilon_1)(\epsilon_2\cdot k_1)$};
\end{scope}
\begin{scope}[xshift=3.8cm,yshift=-3cm]
\coordinate (3) at (-1,0);
\coordinate (1) at (0,0);
\coordinate (2) at (1,0);
\coordinate (4) at (0,1);
\filldraw [particle,color=red,thick ] (3) circle(2pt) node[below=1pt]{$3$}--(1);
\filldraw [particle,color=blue,thick ] (2) circle(2pt) node[below=1pt]{$2$}--(1);
\filldraw [particle,thick] (4) circle(2pt) node[right=1pt]{$4$}--(1) circle(2pt) node[below=1pt]{$1$};
\node at (0,0) [below=0.5cm,font=\fontsize{7}{6}\selectfont]{$n_5=(\epsilon_4\cdot\epsilon_1)(\epsilon_3\cdot k_1)(\epsilon_2\cdot k_1)$};
\end{scope}
\begin{scope}[xshift=7.6cm,yshift=-2.8cm]
\coordinate (3) at (-1,0);
\coordinate (1) at (0,0);
\coordinate (2) at (1,0);
\coordinate (4) at (1,1);
\filldraw (1) circle(2pt) node[below=1pt]{$1$};
\filldraw [particle,color=red,thick ] (3) circle(2pt) node[below=1pt]{$3$}--(1);
\filldraw [particle,thick] (2) circle(2pt) node[below=1pt]{$2$}--(1) ;
\filldraw [particle,thick] (4) circle(2pt) node[right=1pt]{$4$}--(2);
\node at (0,0) [below=0.5cm,font=\fontsize{7}{6}\selectfont]{$n_6=(\epsilon_4\cdot f_2\cdot\epsilon_1)(\epsilon_3\cdot k_1)$};
\end{scope}
\begin{scope}[xshift=-4cm,yshift=-5.6cm]
\coordinate (1) at (-1,.5);
\coordinate (3) at (0,0.5);
\coordinate (2) at (1,0.5);
\filldraw (1) circle(2pt) node[below=4pt]{$1$};
\filldraw[particle] (3) circle(2pt) node[below=4pt]{$3$}--(1);
\filldraw[particle] (2) circle(2pt) node[below=4pt]{$2$}--(3);
\node at (0,0) [below=0.5cm,font=\fontsize{7}{6}\selectfont]{${n}^{\rm YM}_{1\gets3\gets2}=n_7+n_8+n_9$};
\end{scope}
\begin{scope}[yshift=-5.6cm]
\coordinate (1) at (-1,0);
\coordinate (3) at (0,0);
\coordinate (2) at (1,0);
\coordinate (4) at (-1,1);
\filldraw (1) circle(2pt) node[below=1pt]{$1$};
\filldraw [particle,color=blue,thick ] (3) circle(2pt) node[below=1pt]{$3$}--(1);
\filldraw [particle,color=blue,thick ] (2) circle(2pt) node[below=1pt]{$2$}--(3);
\filldraw [particle,thick] (4) circle(2pt) node[right=1pt]{$4$}--(1);
\node at (0,0) [below=0.5cm,font=\fontsize{7}{6}\selectfont]{$n_7=(\epsilon_4\cdot\epsilon_1)(\epsilon_2\cdot f_3 \cdot k_1)$};
\end{scope}
\begin{scope}[xshift=3.8cm,yshift=-5.6cm]
\coordinate (1) at (-1,0);
\coordinate (3) at (0,0);
\coordinate (2) at (1,0);
\coordinate (4) at (0,1);
\filldraw (1) circle(2pt) node[below=1pt]{$1$};
\filldraw [particle,thick] (3) circle(2pt) node[below=1pt]{$3$}--(1) ;
\filldraw [particle,color=blue,thick ] (2) circle(2pt) node[below=1pt]{$2$}--(3);
\filldraw [particle,thick] (4) circle(2pt) node[right=1pt]{$4$}--(3);
\node at (0,0) [below=0.5cm,font=\fontsize{7}{6}\selectfont]{$n_8=(\epsilon_4\cdot f_3 \cdot\epsilon_1)(\epsilon_2\cdot k_3)$};
\end{scope}
\begin{scope}[xshift=7.6cm,yshift=-5.6cm]
\coordinate (1) at (-1,0);
\coordinate (3) at (0,0);
\coordinate (2) at (1,0);
\coordinate (4) at (1,1);
\filldraw [particle,thick] (3)--(1) circle(2pt) node[below=1pt]{$1$} ;
\filldraw (3) circle(2pt) node[below=1pt]{$3$}; 
\filldraw [particle,thick] (2) circle(2pt) node[below=1pt]{$2$}--(3);
\filldraw [particle,thick] (4) circle(2pt) node[right=1pt]{$4$}--(2);
\node at (0,0) [below=0.5cm,font=\fontsize{9}{6}\selectfont]{$n_9=(\epsilon_4\cdot f_2\cdot f_3\cdot\epsilon_1)$};
\end{scope} 
\end{tikzpicture}
\caption{ Cayley numerators of ${\rm Pf}'\Psi_4 $ \label{fig4pt} }
\end{figure}

As usual, with $\s_n \to\infty$ and $1$ as the root, we have $(n-1)^{n{-}3}$ oriented labelled trees, for $n=4$ we have $3$ trees (see \eqref{eq19} and also the leftmost column of Figure~\ref{fig4pt}). There are two major differences compared to ${\rm det}' A_n$ case. First, the Cayley numerator is no longer a monomial of kinematic invariants but rather a sum of $n{-}1$ terms, each associated with an $n$-point tree with $n$ connected to any of the nodes $1,2,\cdots, n{-}1$ (for $n=4$, we list in Figure~\ref{fig4pt} for each of the $3$-point trees on the left, three $4$-point trees all with arrows flowing towards root $1$); second, we need to choose a reference ordering for other labels, $\pi(2) \prec \pi(3) \prec \cdots \prec \pi(n{-}1)$, which determines a decomposition of each $n$-point tree into products of {\it paths}. The first path (called ``baseline''
) 
is the part of the $n$-point tree from node $n$ to node $1$, the second path from node $\pi(2)$ to existing paths (the baseline), the third one from $\pi(3)$ to existing paths, {\it etc.} until the last path from node $\pi(n{-}1)$ to existing paths; if a node has appeared in previous paths, the new path from it is trivial (just the node itself), or alternatively we skip it to consider the next path. In Figure~\ref{fig4pt} we show the (non-trivial) paths for all $9$ trees for the reference ordering $2\prec 3$ with different colors (black for baseline, blue for the path from $2$, and red for the path from $3$ {\it etc.}).

Now the rule is simply that each Cayley numerator is given by the sum of $n{-}1$ monomials, and each of them is the product of the following factors over (non-trivial) paths: for the ``baseline'' $n \to j \cdots i \to 1$ we assign a factor $\epsilon_n \cdot f_j \cdots f_i \cdot \epsilon_1$ and for any other (non-trivial) path $a \to b \cdots \to c$, we assign $\epsilon_a \cdot f_b \cdots k_c$~\footnote{Here $f_i^{\mu \nu}$ is the linearized field-strength $f_i^{\mu \nu}=k^\mu_i \epsilon^\nu_i-k^\nu_i \epsilon^\mu_i$  and we contract the Lorentz indices according to the path. Note that we do not expand $f$ but regard such a contraction as a ``monomial'' of kinematic invariants.}. For $n=4$, we have listed the $9$ monomials for the $9$ trees, and we have ${n}^{\rm YM}_{1\gets2\gets3}=n_1+n_2+n_3$, ${n}^{\rm YM}_{3\to1\gets2}=n_4+n_5+n_6$ and ${n}^{\rm YM}_{1\gets3\gets2}=n_7+n_8+n_9$. From here it is straightforward to obtain kinematic numerators by \eqref{ng} for four-point amplitude in YM and (by squaring) those in GR, and by using numerators \eqref{eq25} of NLSM we obtain four-point numerators for BI amplitude:
\be
n^{\rm BI}_s\!=\!-s^2 ({n}^{\rm YM}_{1\gets2\gets3}+{n}^{\rm YM}_{3\to1\gets2}), ~\, n^{\rm BI}_t\!=\!t(s-u)({n}^{\rm YM}_{1\gets2\gets3}-{n}^{\rm YM}_{1\gets3\gets2}),~\, n^{\rm BI}_u\!=\!-u^2({n}^{\rm YM}_{1\gets3\gets2}+{n}^{\rm YM}_{3\to1\gets2})\,.
\ee

Note that unlike the case for NLSM and sGal, here the symmetry among $2,3,\cdots, n{-}1$ is broken by our choice of reference ordering $\rho$. 
It can be recovered by manually averaging over all $(n{-}2)!$ $\rho$'s, or by using the implicitly symmetrized construction of \cite{Edison:2020ehu}.

Although this ``symmetrized'' version gives longer expressions for ${\bf n}_\Gamma$, it has the advantage that for computing amplitudes, only a small fraction of all cubic trees are needed (all others given by relabelling), just as what we did for NLSM and sGal. For example, for $8$-point GR, the symmetrized version needs only $46$ out of $11!!$ cubic trees, increasing the efficiency by more than $200$ times. 

In the {\sc Mathematica} package, we provide functions 
\ba 
&{\texttt {AmpGR}}[ n\_], \qquad\qquad\quad\qquad\quad
{\texttt {nCubicGR}}[poles\_] ,
\nl
&{\texttt {AmpBI}}[ n\_], \qquad\qquad\quad\qquad\quad
{\texttt {nCubicBI}}[poles\_] ,
\nl 
&{\texttt {AmpYM}}[ ordering\_], \qquad\qquad\quad {\texttt {nCubicYM}}[poles\_, ordering\_] ,
\nonumber
\ea 
 in the {\sc Mathematica} package that generate the amplitudes or numerators of cubic trees  of related theories
 and 
 \ba 
{\texttt {nCubicYM}}[poles\_], \qquad {\texttt {nCayleyYM}}[pairs\_\_] ,
 \ea 
 for numerators of cubic trees or labelled trees of ${\rm Pf}{}'\Psi_n$.  For example, one can get an $n=5$ Cayley numerator, say $n\left(
\raisebox{-10pt}{
\tikz[scale=.5]{	\filldraw (0,0) circle(2pt)  node[below=1.5pt]{$2$};
\filldraw (1,0) circle(2pt)  node[below=1.5pt]{$1$};
\filldraw (2,0) circle(2pt)  node[below=1.5pt]{$3$};
\filldraw (3,0) circle(2pt)  node[below=1.5pt]{$4$};
	\draw[particle] (0,0)--(1,0);
	\draw[particle] (2,0)--(1,0);
		\draw[particle] (3,0)--(2,0);
}
} \right)$
, via the package using
\ba
\texttt {nCayleyYM}[\{2,1\}, \{3,1\}, &\{4,3\} ]=\,(\e_5\cdot f_2\cdot \e_1)(\e_3\cdot k_1)(\e_4\cdot k_3)+(\e_5\cdot \e_1)(\e_2\cdot k_1)(\e_3\cdot k_1)(\e_4\cdot k_3)\nonumber\\
&+(\e_5\cdot f_3\cdot \e_1)(\e_2\cdot k_1)(\e_4\cdot k_3)+(\e_5\cdot f_4\cdot f_3\cdot\e_1)(\e_2\cdot k_1).
\ea 
A cubic tree numerator of partial YM amplitudes can be generated via $\texttt {nCubicYM}$, {\it e.g.}

\def\Cdot{\!\cdot\!}
{\footnotesize{
\ba
&\texttt{nCubicYM}[s_{1,2}s_{3,4},\{1,2,3,4,5\}]
\\
=&-(\!\e_2\Cdot k_1 \!)(\!\e_3\Cdot k_2\!)(\! \e_4\Cdot k_3 \!)(\!\e_5\Cdot \e_1\!)
-(\!\e_3\Cdot k_2 \!)(\!\e_4\Cdot k_3\!)(\! \e_5\Cdot f_2\Cdot \e_1\!)
-(\!\e_4\Cdot k_3\!)(\! \e_5\Cdot f_3\Cdot f_2\Cdot \e_1\!)
-(\!\e_5\Cdot f_4\Cdot f_3\Cdot f_2\Cdot \e_1\!)
\nl&
-(\!\e_2\Cdot k_1 \!)(\!\e_3\Cdot k_1 \!)(\!\e_4\Cdot k_3 \!)(\!\e_5\Cdot \e_1 \!)
-(\!\e_3\Cdot k_1 \!)(\!\e_4\Cdot k_3 \!)(\! \e_5\Cdot f_2\Cdot \e_1 \!)
-(\!\e_2\Cdot k_1 \!)(\! \e_4\Cdot k_3 \!)(\!\e_5\Cdot f_3\Cdot \e_1 \!)
-(\!\e_2\Cdot k_1 \!)(\! \e_5\Cdot f_4\Cdot f_3\Cdot \e_1 \!)
\nl&
+(\!\e_2\Cdot k_1 \!)(\!\e_3\Cdot f_4\Cdot k_1 \!)(\!\e_5\Cdot \e_1\!)
+ (\!\e_3\Cdot f_4\Cdot k_1\!)(\! \e_5\Cdot f_2\Cdot \e_1\!)
+ (\!\e_2\Cdot k_1\!)(\! \e_5\Cdot f_4\Cdot \e_1\!)(\!\e_3\Cdot k_4 \!)
+ (\!\e_2\Cdot k_1\!)(\! \e_5\Cdot f_3\Cdot f_4\Cdot \e_1\!)
\nl&
+(\!\e_2\Cdot k_1\!)(\! \e_3\Cdot f_4\Cdot k_2\!)(\! \e_5\Cdot \e_1\!)
+(\!\e_3\Cdot f_4\Cdot k_2\!)(\! \e_5\Cdot f_2\Cdot \e_1\!)
+(\!\e_5\Cdot f_4\Cdot f_2\Cdot \e_1\!)(\!\e_3\Cdot k_4 \!)
+(\!\e_5\Cdot f_3\Cdot f_4\Cdot f_2\Cdot \e_1\!)\,,
\nonumber
\ea}}
where each line on the RHS is a Cayley numerator for a labelled tree (see \eqref{5ptex2}). In this way we can generate amplitudes in YM, GR and BI easily. With $\texttt {AmpGR}[ n, $ $ \textsl {Symmetric} \to \textsl {True}]$ and  $\texttt {AmpBI}[ n, \textsl {Symmetric} \to \textsl {True}]$ the code computes GR and BI amplitudes really fast for {\it e.g.} $n=7,8$, which, as far as we know, is rather time-consuming using other methods.

Since the Cayley numerators from ${\rm Pf}' \Psi_n$ are rather complicated, let us now move to a simplified setting where they can be written more explicitly. We consider dimension reduction of YM to YMs and keep all (even $n$) external legs as scalars (thus all Cayley numerators will be Mandelstam variables), and without loss of generality we consider scalar pairs {\it e.g.} $(n1)(23) \cdots (n{-}2\quad n{-}1)$. In other words, we expand the half-integrand ${\rm Pf'} A_n/(\s_{n1} \s_{23} \cdots \s_{n{-}2 ,n{-}1})$, which amounts to kinematic numerators for all scalar amplitudes in YMs, EMs and DBI (together with those from ${\rm det}' A_n$). 
It turns out that only those labelled trees that have edges $n\dedge1$, $2\dedge3$, $\cdots, (n{-}2)\dedge(n{-}1)$ can contribute! For this reason, we first draw all the ``skeleton'' labelled trees by viewing each pair $(i{-}1\quad i)$ as a node, and we denote these $n/2$ nodes by $i=1, 3, \cdots, n{-}1$. We further choose a reference ordering for these nodes {\it e.g.} $1 \prec 3 \prec \cdots \prec n{-}1$, and similar to the rule for YM, there is a decomposition into ``paths'': the first path is trivial (node $1$ itself), the second from node $3$ to 
an existing path (node $1$), and the third from node $5$ to existing paths, {\it etc.}. Below we list the $3$ skeleton trees for $n=6$ example with scalar pairs $(61)(23)(45)$, and show the paths using different colors (black and blue),

\ba
\raisebox{-15pt}{
\tikz{	\filldraw (0,0) circle(2pt)  node[below=1.5pt]{$(61)$};
\filldraw (1,0) circle(2pt)  node[below=1.5pt]{$(45)$};
\filldraw (2,0) circle(2pt)  node[below=1.5pt]{$(23)$};
	\draw[particle,thick] (1,0)--(0,0);
	\draw[particle,thick] (2,0)--(1,0);
	}
}\,,\quad 
\raisebox{-15pt}{
\tikz{	\filldraw[color=blue] (0,0) circle(2pt)  node[below=1.5pt,black]{$(45)$};
\filldraw (1,0) circle(2pt)  node[below=1.5pt]{$(61)$};
\filldraw (2,0) circle(2pt)  node[below=1.5pt]{$(23)$};
	\draw[particle,color=blue,thick] (0,0)--(1,0);
	\draw[particle,thick] (2,0)--(1,0);
}
}\,,\quad 
\raisebox{-15pt}{
\tikz{	\filldraw (0,0) circle(2pt)  node[below=1.5pt]{$(61)$};
\filldraw (1,0) circle(2pt)  node[below=1.5pt]{$(23)$};
\filldraw[color=blue] (2,0) circle(2pt)  node[below=1.5pt,black]{$(45)$};
	\draw[particle,thick] (1,0)--(0,0);
	\draw[particle,color=blue,thick] (2,0)--(1,0);
}}
\,.
\ea 

To get Cayley functions, we still need to ``blow up'' each node into an edge (except for node $1=(n 1)$ since $\s_n \to\infty$): for any node that is not a starting point of a (non-trivial) path we can have a labelled tree with the edge from $i{-}1$ to $i$ or from $i$ to $i{-}1$, but for a node that is the starting point  we further require $i{-}1 \prec i$. We list the $5$ labelled trees after blowing up for our $n=6$ example, 
\ba
\label{eq317}
\raisebox{-20pt}{
\begin{tikzpicture}[neuron/.style = {circle, minimum size=3pt, inner sep=0pt,  fill=black } ]
		\draw[dashed] (4.3,-.5)--(4.3,1.2);
		\draw[dashed] (6.8,-.5)--(6.8,1.2);
		    \begin{scope}[xshift=-1.2cm]
				\node[neuron] (1) at (-1.4,0) {} ;
				\node[neuron] (2) at (-.7,0) {} ;
				\node[neuron] (3) at (0,0) {} ;
				\node[neuron] (4) at (.7,0) {} ;
				\node[neuron] (5) at (1.4,0) {} ;
				\draw [particle] (2) node[below=1pt]{$4$}--(1) node[below=1pt]{$1$};
				\draw [particles] (3) node[below=1pt]{$5$}--(2);
				\draw [particle] (4) node[below=1pt]{$3$}--(3);
				\draw [particles] (5) node[below=1pt]{$2$}--(4);
			\end{scope}	
			\begin{scope}[xshift=2.4cm]
				\node[neuron] (1) at (-1.4,0) {} ;
				\node[neuron] (2) at (-.7,0) {} ;
				\node[neuron] (3) at (0,0) {} ;
				\node[neuron] (4) at (.7,0) {} ;
				\node[neuron] (5) at (1.4,0) {} ;
				\draw [particle] (2) node[below=1pt]{$5$}--(1) node[below=1pt]{$1$};
				\draw [particles] (3) node[below=1pt]{$4$}--(2);
				\draw [particle] (4) node[below=1pt]{$3$}--(3);
				\draw [particles] (5) node[below=1pt]{$2$}--(4);
			\end{scope}	
		    \begin{scope}[xshift=5.5cm]
		        \node[neuron] (1) at (-.8,0) {} ;
		        \node[neuron] (2) at (0,0) {} ;
				\node[neuron] (3) at (.8,0) {} ;
				\node[neuron] (4) at (-.4,.8) {} ;
				\node[neuron] (5) at (.4,.8) {} ;
				\filldraw [particle] (2) node[below=1pt]{$3$}--(1) node[below=1pt]{$1$};
				\filldraw [particles] (3) node[below=1pt]{$2$}--(2);
				\filldraw [particles] (5) node[above=1pt]{$4$}--(4);
				\filldraw [particle] (4)node[above=1pt]{$5$}--(1);
			\end{scope}	
			\begin{scope}[xshift=8cm]
				\node[neuron] (1) at (-.8,0) {} ;
				\node[neuron] (2) at (0,0) {} ;
				\node[neuron] (3) at (.8,0) {} ;
				\node[neuron] (4) at (-.4,.8) {} ;
				\node[neuron] (5) at (.4,.8) {} ;
				\filldraw [particle] (2) node[below=1pt]{$3$}--(1) node[below=1pt]{$1$};
				\filldraw [particles] (3) node[below=1pt]{$2$}--(2);
				\filldraw [particles] (5) node[above=1pt]{$4$}--(4);
				\filldraw [particle] (4)node[above=1pt]{$5$}--(2);
			\end{scope}	
			\begin{scope}[xshift=10.5cm]
				\node[neuron] (1) at (-.8,0) {} ;
				\node[neuron] (2) at (0,0) {} ;
				\node[neuron] (3) at (.8,0) {} ;
				\node[neuron] (4) at (-.4,.8) {} ;
				\node[neuron] (5) at (.4,.8) {} ;
				\filldraw [particle] (2) node[below=1pt]{$3$}--(1) node[below=1pt]{$1$};
				\filldraw [particles] (3) node[below=1pt]{$2$}--(2);
				\filldraw [particles] (5) node[above=1pt]{$4$}--(4);
				\filldraw [particle] (4)node[above=1pt]{$5$}--(3);
			\end{scope}		
		\end{tikzpicture}}
\ea
where solid(dashed) lines are used for old(new) edges. 
The upshot is that for such a labelled tree $\Gamma$, ${ n}_\Gamma$ is given by a product of $s_{i j}$ for the $\frac n 2-1$ solid edges.  The Cayley expansion of the half-integrand reads 
\be
\boxed{
\sum_{\Gamma}
C_\Gamma \left( \prod_{(ij) \in E_{\rm solid} (\Gamma)}^{\frac n 2-1} s_{i j} \right)\,,
}
\ee
where we sum over all labelled trees after blowing up as above. For each of them, the Cayley numerator ${ n}_\Gamma$ is given by the product of $s_{ij}$'s for the $\frac n 2-1$ ``solid'' edges as shown in the parenthesis; equivalently it is given by the product over all edges, divided by the $n/2$ factors for ``dashed'' edges $s_{n1} s_{23} \cdots s_{n{-}2, n{-}1}$.
For our $n=6$ example, the result from the above $5$ trees reads
\be\label{eq319}
\frac{s_{14} s_{35}}{\s_{23} \s_{35}\s_{54} \s_{41}  }
+\frac{s_{15} s_{34}}{\s_{23} \s_{34}\s_{45}\s_{51}     }
+\frac{s_{13} s_{15}}{\s_{45}\s_{51}\s_{23}\s_{31} }
+\frac{s_{13} s_{35}}{\s_{45} \s_{53} \s_{23} \s_{31}}
+\frac{s_{13} s_{25}}{\s_{45} \s_{52} \s_{23} \s_{31}} \,.
\ee

Similarly, in the {\sc Mathematica} package we have functions 
 \ba \label{yms1}
{\texttt {nCubicYMs}}[poles\_,particles\_\_], \qquad {\texttt {nCayleyYMs}}[\{pairs\_\_\},particles\_\_] ,
 \ea 
that generate numerators for cubic trees or labelled trees from ${\rm Pf}X_n{\rm Pf}{}'A_n$. For example, one can get the Cayley numerator of the first labelled tree in \eqref{eq317} via 
 \ba 
 {\texttt {nCayleyYMs}}[\{\{4,1\},\{5,4\},\{3,5\},\{2,3\}\},\{2,3\},\{4,5\},\{1,6\}]=s_{14}s_{35}\,,
 \ea 
 where $\{\{4,1\},\{5,4\},\{3,5\},\{2,3\}\}$ specifies (edges of) the labelled tree and $\{6,1\}$,$\{2,3\}$,$\{4,5\}$ indicates different traces for the particles. Following \eqref{eq319}, we can obtain kinematic numerators for any cubic trees according to \eqref{ng}, which are implemented by the function $ {\texttt {nCubicYMs}}$. For example, the numerator of the cubic tree \raisebox{-15pt}{
\begin{tikzpicture}
\node (1) at (-1.5,0) {};
\node (2) at (-1,0.6) {};
\node (3) at  (-.5,0.6) {};
\node (4) at (0,0.6) {};
\node (5) at (0.5,0.6) {};
\node (6) at (1,0) {};
\draw (1)  node[left=1pt]{$2$} --(6);	
\draw (2) node[above=-2pt]{$3$}--(-1,0);	
\draw (3) node[above=-2pt]{$4$}--(-.5,0);
\draw (4) node[above=-2pt]{$5$}--(0,0);
\draw (5) node[above=-2pt]{$6$}--(0.5,0);
\draw (6) node[right=1pt]{$1$};
\end{tikzpicture}}
(which in the code is denoted as $s_{23}s_{234}s_{2345}$) is given by
\ba \label{eq322}
 {\texttt {nCubicYMs}}[s_{2,3}s_{2,3,4}s_{2,3,4,5},\{2,3\},\{4,5\},\{1,6\}]= \mp  s_{34}s_{15}\,,
\ea
 where the sign $\mp$ is given explicitly in the code but it does not matter here.  
 
Amplitudes or cubic tree numerators of related theories (YMs, EMs and DBI) are implemented by the following functions
 \ba \label{yms2}
&{\texttt {AmpEYM}}[ particles\_\_], \qquad\qquad\quad\qquad\quad
{\texttt {nCubicEYM}}[poles\_,particles\_\_] ,
\nl
&{\texttt {AmpDBI}}[particles\_\_], \qquad\qquad\quad\qquad\quad
{\texttt {nCubicDBI}}[poles\_,particles\_\_] ,
\nl 
&{\texttt {AmpYMs}}[particles\_\_, ordering\_], \qquad\quad {\texttt {nCubicYMs}}[poles\_, particles\_\_, ordering\_] .
\nonumber
 \ea 
Again,  $ {\texttt {nCubicYMs}}[s_{2,3}s_{2,3,4}s_{2,3,4,5},\{2,3\},\{4,5\},\{1,6\},\{1,2,3,4,5,6\}]$ gives the same result for the planar tree as $ {\texttt {nCubicYMs}}[s_{2,3}s_{2,3,4}s_{2,3,4,5},
$ $\{2,3\},\{4,5\},\{1,6\}]$ in \eqref{eq322}, except that the sign $\pm$ is now fixed as $-$. Considering numerators of  all 14 planar cubic trees, we obtain the color-ordered YMs amplitude for this trace structure:
\ba 
&{\texttt {AmpYMs}}[\{2,3\},\{4,5\},\{1,6\},\{1,2,3,4,5,6\}]
\\
=&-\frac{s_{1 5}}{s_{1 6} s_{2 3 4}}-\frac{s_{3 4} s_{1 5}}{s_{1 6} s_{2 3} s_{2 3 4}}-\frac{s_{1 5}}{s_{1 6} s_{3 4 5}}+\frac{-s_{1 5} s_{3 4}+s_{1 4} s_{3 5}+s_{1 3} \left(s_{2 5}+s_{3 5}\right)}{s_{1 6} s_{2 3} s_{4 5}}
\nl 
\nonumber
&+\frac{s_{1 3} \left(s_{1 5}+s_{2 5}+s_{3 5}\right)}{s_{2 3} s_{4 5} s_{1 2 3}}+\frac{\left(s_{1 3}+s_{1 4}\right) s_{3 5}-s_{1 5} s_{3 4}}{s_{1 6} s_{4 5} s_{3 4 5}}
\, .
\ea 
As mentioned, our {\sc Mathematica} package applies to general (mixed) amplitudes in gen. YMs and EYM. The functions given in \eqref{yms1} and \eqref{yms2} apply to cases where there are more than two particles in a trace. For instance, a color-ordered double-trace amplitude with 3 bi-adjoint scalars in each trace reads
\ba 
&{\texttt {AmpYMs}}[\{1,2,5\},\{3,4,6\},\{1,2,3,4,5,6\}]
\\
\nonumber
=&\frac{s_{2 3}+s_{2 4}}{s_{1 6} s_{3 4} s_{2 3 4}}+\frac{1}{s_{1 6} s_{2 3 4}}-\frac{1}{s_{1 2} s_{3 4 5}}-\frac{1}{s_{1 6} s_{3 4 5}}-\frac{s_{3 5}+s_{4 5}}{s_{1 2} s_{3 4} s_{3 4 5}}-\frac{s_{3 5}+s_{4 5}}{s_{1 6} s_{3 4} s_{3 4 5}}
\,.
\ea 
Another example is the scattering of two gravitons $1_h,2_h$ and two gluons $3_g,4_g$ in a trace, the $t$-channel cubic tree numerator reads
 \ba 
&{\texttt {nCubicEYM}}[s_{2,3},\{1\},\{2\},\{3,4\}]
\\
=& 
\Big( \tilde\epsilon _1\cdot \tilde {f}_2\cdot k_3+ (\tilde\epsilon _1\cdot k_3 )( \tilde\epsilon _2\cdot k_3) \Big)
 \Big( 
 (\epsilon _1\cdot \epsilon _4)( \epsilon _2\cdot f_3\cdot k_1)-(\epsilon _4\cdot f_2\cdot \epsilon _1)( k_2\cdot \epsilon _3)
 \nl 
 &+(\epsilon _4\cdot f_3\cdot \epsilon _1)( k_3\cdot \epsilon _2)+(\epsilon _4\cdot f_2\cdot f_3\cdot \epsilon _1)-(\epsilon _4\cdot f_3\cdot f_2\cdot \epsilon _1)-(\epsilon _1\cdot \epsilon _4) (k_1\cdot \epsilon _2) (k_2\cdot \epsilon _3)
 \Big) 
 \,,
\nonumber
 \ea 
where we use $\{i\}$ for a graviton, and the full amplitude is given by $ {\texttt {AmpEYM}}[\{1\},\{2\},\{3,4\}]$. 
 Again, we remind the reader to use $?$ to learn about any {\sc Mathematica} function appeared in the paper (see \eqref{inquiring}). 

Before ending, we remark that these numerators for YMs amplitudes with $n$ pairs of scalars as obtained above also give explicit results for YM $n$-point amplitudes in the form of ``scalar-scaffolded gluons''~\cite{Arkani-Hamed:2023jry, Arkani-Hamed:2023swr} and via double copy also new results for ``scaffolded'' gravity amplitudes. It would be interesting to investigate further along this line. 
\section{Outlook}

The main result of this  note is the formula \eqref{ng} for expressing the kinematic numerator of any cubic tree in terms of Cayley numerators of labelled trees, which can be viewed as basic building blocks for kinematic numerators. As we have seen, each Cayley numerator is either a monomial of kinematic invariants, or for ${\rm Pf}'\Psi_n$ it is a sum of $n{-}1$ such terms, and our formula gives all kinematic numerators of all known amplitudes admitting CHY representation as linear combinations of such building blocks. 

Our {\sc Mathematica} code implementing this method seems to be quite efficient for computing tree amplitudes: although it involves a large number of labelled trees, it avoids the time-consuming shuffle computation for master numerators and that for the $(n{-}2)!$-dim matrix $m(\alpha|\beta)$ in the usual representation; the symmetrized version (automatic for ${\rm det}'A_n$) increases the efficiency greatly, and it would be interesting to see if we can further improve it. More importantly, our result for the kinematic numerators in terms of simple building blocks may shed new lights into {\it kinematic algebras} for various theories ({\it c.f.}  \cite{Chen:2019ywi, Cheung:2016prv}). In this regard, we note that for NLSM they are rather symmetric and one can nicely see Adler zero in the expansion, but for YM {\it etc.} the choice of reference ordering {\it etc.} obscures the symmetry and possible underlying structure, where it becomes difficult to see gauge invariance manifestly (which is essentially the defining property here, see ~\cite{Arkani-Hamed:2016rak, Rodina:2020jlw} ). Last but not least, our method is closely related to that in \cite{Frost:2020eoa} (see also \cite{Frost:2019fjn, Mafra:2020qst}), and it would be interesting to work out the connections more precisely.

As mentioned, our tree-level result can be uplifted to one loop via forward limit with a pair of momenta in higher dimensions (see \cite{He:2017spx, Edison:2020uzf}), which gives integrand with propagators linear in loop momentum. It would be highly desirable to find a systematic method for extracting kinematic numerators efficiently at loop level with quadratic propagators (see~\cite{Bridges:2021ebs},\cite{Agerskov:2019ryp} and references therein). Finally, our original motivation for introducing Cayley functions is due to their geometric interpretation, {\it e.g.} each $C_\Gamma$ is mapped via scattering equations to the canonical form of a polytope in kinematic space~\cite{Arkani-Hamed:2017mur}; it would be fascinating to see if our way of expanding YM/NLSM amplitudes by numerators leads to a geometric interpretation to {\it scattering forms} of gluons and pions.

	
\section*{Acknowledgement}
We would like to thank Alexander Edison and Carlos Mafra for valuable comments. The research of S. H. is supported in part by National Natural Science Foundation of China under Grant No. 11935013,11947301, 12047502,12047503. The research of Y.Z. is supported by the Knut and Alice Wallenberg Foundation under the grant KAW 2018.0116: From Scattering Amplitudes to Gravitational Waves.
	\appendix

\section{Derivation of  ${\rm sgn}^\rho_\Gamma$ }\label{signsign}
	In this appendix, we derive the relative signs used in \eqref{ng}. These signs follow those in \eqref{cayleysign} so we prove them at first.  
	As proved in  \cite{Gao:2017dek},  the overall sign of $m(\Gamma|\Gamma')$  in \eqref{cayleysign} is the same as that of $m(\rho|\rho')$,
	\ba \label{a1}
	\<g|\Gamma\> \<g|\Gamma'\>=
		\<g|\rho\> \<g|\rho'\>
		\ea 
	where $g$ could be any cubic tree belonging to $T(\Gamma)\cap T(\Gamma')$  and  $\rho,\rho'$ are two planar orderings of $g$, {\it i.e.} two Hamilton graphs, which are given by 
	\ba \label{a2}
	\rho={\cal O}^g_{\Gamma}[1,2,\cdots,n-1]\,,\quad
	\rho'={\cal O}^g_{\Gamma'}[1,2,\cdots,n-1]\,.
	\ea 
 The operation ${\cal O}$ is originally defined in \cite{Gao:2017dek} which maps a set of nodes to an {\it ordered} sequence.  
	The superscript and subscript imply that the operation ${\cal O}^g_{\Gamma}$ depends on the structure of the cubic tree $g$ and the labelled tree $\Gamma$.  The operation can be performed recursively. Recall that there must be an edge $i-j$ in $\Gamma$ if there is a pole $s_{ij}$ in $g$ according to \eqref{pc} and we define
	\ba \label{a3}
	{\cal O}^g_{\Gamma}[i,j]=\begin{cases}
	[i,j] \quad {\text{if the orientation in $\Gamma$ is $i\to j$}}
	\\
		[j,i] \quad {\text{if the orientation in $\Gamma$ is $j\to i$}}
	\end{cases}
	\,.
	\ea 
	As already explained in sec \ref{secFg}, 
whenever there is a multi-particle pole $s_I$ in $g$, there must be  either 
two poles $s_{I_1},s_{I_2}$ with $I_1\sqcup I_2=I$, or
a pole with exactly one less point, $s_{I/\{i\}}$.  In the first case, there must be an edge $i-j$ linking 	two subgraphs with nodes $I_1\ni i$ and $I_2 \ni j$ respectively and we define
\ba\label{a4} 
	{\cal O}^g_{\Gamma}[I_1,\,I_2]=\begin{cases}
	\big[{\cal O}^g_{\Gamma}[I_1],\,{\cal O}^g_{\Gamma}[I_2] \big] \quad {\text{if the orientation in $\Gamma$ is $i\to j$ with $i\in I_1, j\in I_2$}}
	\\
		\big[{\cal O}^g_{\Gamma}[I_2],\,{\cal O}^g_{\Gamma}[I_1] \big] \quad {\text{if the orientation in $\Gamma$ is $j\to i$ with $i\in I_1, j\in I_2$}}
	\end{cases}
	\,.
	\ea 
Similarly, in the second case,
there must be an edge $i-j$ linking the note $i$ to another subgraph with nodes $I/\{i\}$ and we define
	\ba \label{a5}
	{\cal O}^g_{\Gamma}[i,\,I/\{i\}]=\begin{cases}
	\big[i,\,{\cal O}^g_{\Gamma}[I/\{i\}] \big] \quad {\text{if the orientation in $\Gamma$ is $i\to j$ with $j\in I/\{i\}$}}
	\\
		[{\cal O}^g_{\Gamma}[I/\{i\}],\,i] \quad {\text{if the orientation in $\Gamma$ is $j\to i$ with $j\in I/\{i\}$}}
	\end{cases}
	\,.
	\ea 
	As has already been used in sec \ref{secFg},  
	we adopt the trick to  introduce an additional pole $s_{1,2,\cdots,n-1}$
	in addition to the $n-3$ poles of $g$ such that 
	we can apply \eqref{a3}-\eqref{a5} to ${\cal O}^g_{\Gamma}[1,2,\cdots,n-1]$ and ${\cal O}^g_{\Gamma'}[1,2,\cdots,n-1]$ and recursively we get the final ordering $\rho$ and $\rho'$. 
	As pointed out in \cite{Mafra:2016ltu},  the overall sign of usual biadjoint amplitude $m(\rho|\rho')$ is nothing but
	\ba 
	\<g|\rho\> \<g|\rho'\>={\rm sgn}^\rho_{\rho'}\,,
	\ea 
where $g\in T(\rho)\cap T(\rho')$ and	$	{\rm sgn}^\rho_{\rho'}= 	{\rm sgn}^{\rho'}_{\rho}$ is given in \eqref{sign}.  Now we propose that 
\ba \label{a7}
\<g|\Gamma\> \<g|\Gamma'\> ={\rm sgn}^{\rho}_{\Gamma} {\rm sgn}^{\rho}_{\Gamma'} \,.
\ea 
\begin{proof}
Note that by construction we have ${\rm sgn}^\rho_{\Gamma}={\rm sgn}^{\rho'}_{\Gamma'}=1$. That is we just need to prove ${\rm sgn}^{\rho}_{\Gamma'}= {\rm sgn}^{\rho}_{\rho'}$, which is also almost confirmed by the construction rules \eqref{a3}-\eqref{a5}. Consider
every pole $s_I$ of $g$ including the fake one $s_{1,2,\cdots,n-1}$. On the one hand, there must be an oriented edge $i\to j$ in the labelled tree $\Gamma'$ linking two subgraphs with nodes $I_1 \ni i$ and $I_2 \ni j$  respectively  (here we allow $|I_1|,|I_2|\geq 1$ to unify the discussion).  On the other hand, regarding $\rho'$ as a Hamilton  graph,  there must be an oriented edge $i'\to j'$ in the labelled tree $\rho'$ linking two subgraphs with nodes $I_1 \ni i'$ and $I_2 \ni j'$  respectively. This shows how $n-2$ edges of $\Gamma'$ are related to those of $\rho'$. For each pair of edges $i\to j$ and $i'\to j'$,  there must also be two disconnected subgraphs in the labelled tree $\rho$ with nodes $I_1 \supset \{i,i'\}$ and $I_2 \supset \{j,j'\}$ respectively, which directly gives
\ba 
{\rm sgn}^{\rho}_{i,j} ={\rm sgn}^{\rho}_{i',j'}\,.
\ea 
Hence according to \eqref{sign}, we proved ${\rm sgn}^{\rho}_{\Gamma'}= {\rm sgn}^{\rho}_{\rho'}$, {\it i.e.}, \eqref{a7} at least holds for the planar ordering $\rho$ defined in \eqref{a2} which strongly depends on the structure of the cubic tree $g$  and the orientations of edges of the labelled tree $\Gamma$.

If we just flip the  orientation of one edge of the labelled tree ${\Gamma}$ and denote the new one as ${\tilde \Gamma}$, we will get $C_{\tilde \Gamma}=-C_\Gamma$ and another identity similar to \eqref{a7},
 \ba 
 \<g|\tilde\Gamma\> \<g|\Gamma'\> ={\rm sgn}^{\tilde\rho}_{\tilde\Gamma} {\rm sgn}^{\tilde\rho}_{\Gamma'} \,,
 \ea 
 where $\tilde\rho={\cal O}^g_{\tilde\Gamma}[1,2,\cdots,n-1]$ is another planar ordering of the cubic tree 
 $g$ in addition to $\rho$ defined in \eqref{a2}.  Owing to $\<g|\tilde\Gamma\>=-\<g|\Gamma\>$ and ${\rm sgn}^{\tilde\rho}_{\tilde\Gamma}= - {\rm sgn}^{\tilde\rho}_{\Gamma} $, the above identity is nothing but the original one \eqref{a7} with $\rho\to \tilde \rho$. 
Considering all possible flips of orientations of $\Gamma$ or $\Gamma'$, one can prove that \eqref{a7} holds for any planar ordering $\rho$ of $g$ including the one defined in \eqref{a2}. 
\end{proof}

So in general we have
	\ba
	\int \dif \mu_n C_\Gamma C_{\Gamma'}=\< \Gamma|\Gamma'\>  =  {\rm sgn}^{\rho}_{\Gamma} {\rm sgn}^{\rho}_{\Gamma'}   \sum_{g\in  T(\Gamma)\cap T(\Gamma') }    \< g|g \>\,,
	\ea	
where $\rho$ could be any planar ordering of any $g\in  T(\Gamma)\cap T(\Gamma') $. 
Since $\<g|\Gamma'\>$ is just a sign, \eqref{a7} implies that 
\ba 
\<g|\Gamma\>  = \<g|\Gamma'\> {\rm sgn}^{\rho}_{\Gamma} {\rm sgn}^{\rho}_{\Gamma'} \,.
\ea 
In particular, when $\Gamma'=\rho$, we have 
\ba
\<g|\Gamma\> = \<g|\rho\> {\rm sgn}^{\rho}_{\Gamma} \,.
\ea 
Together with \eqref{ng1}, it	 leads to \eqref{ng}.

\bibliographystyle{JHEP}
\bibliography{Refs}

\providecommand{\href}[2]{#2}\begingroup\raggedright\begin{thebibliography}{10}

\bibitem{Bern:2008qj}
Z.~Bern, J.~J.~M. Carrasco, and H.~Johansson, {\it {New Relations for
  Gauge-Theory Amplitudes}},  {\em Phys. Rev. D} {\bf 78} (2008) 085011,
  [\href{http://arxiv.org/abs/0805.3993}{{\tt arXiv:0805.3993}}].

\bibitem{Bern:2010ue}
Z.~Bern, J.~J.~M. Carrasco, and H.~Johansson, {\it {Perturbative Quantum
  Gravity as a Double Copy of Gauge Theory}},  {\em Phys. Rev. Lett.} {\bf 105}
  (2010) 061602, [\href{http://arxiv.org/abs/1004.0476}{{\tt
  arXiv:1004.0476}}].

\bibitem{Bern:2017yxu}
Z.~Bern, J.~J. Carrasco, W.-M. Chen, H.~Johansson, and R.~Roiban, {\it {Gravity
  Amplitudes as Generalized Double Copies of Gauge-Theory Amplitudes}},  {\em
  Phys. Rev. Lett.} {\bf 118} (2017), no.~18 181602,
  [\href{http://arxiv.org/abs/1701.02519}{{\tt arXiv:1701.02519}}].

\bibitem{Bern:2019prr}
Z.~Bern, J.~J. Carrasco, M.~Chiodaroli, H.~Johansson, and R.~Roiban, {\it {The
  Duality Between Color and Kinematics and its Applications}},
  \href{http://arxiv.org/abs/1909.01358}{{\tt arXiv:1909.01358}}.

\bibitem{Bern:2010yg}
Z.~Bern, T.~Dennen, Y.-t. Huang, and M.~Kiermaier, {\it {Gravity as the Square
  of Gauge Theory}},  {\em Phys. Rev. D} {\bf 82} (2010) 065003,
  [\href{http://arxiv.org/abs/1004.0693}{{\tt arXiv:1004.0693}}].

\bibitem{Bern:2012uf}
Z.~Bern, J.~J.~M. Carrasco, L.~J. Dixon, H.~Johansson, and R.~Roiban, {\it
  {Simplifying Multiloop Integrands and Ultraviolet Divergences of Gauge Theory
  and Gravity Amplitudes}},  {\em Phys. Rev. D} {\bf 85} (2012) 105014,
  [\href{http://arxiv.org/abs/1201.5366}{{\tt arXiv:1201.5366}}].

\bibitem{Bern:2012cd}
Z.~Bern, S.~Davies, T.~Dennen, and Y.-t. Huang, {\it {Absence of Three-Loop
  Four-Point Divergences in N=4 Supergravity}},  {\em Phys. Rev. Lett.} {\bf
  108} (2012) 201301, [\href{http://arxiv.org/abs/1202.3423}{{\tt
  arXiv:1202.3423}}].

\bibitem{Bern:2013uka}
Z.~Bern, S.~Davies, T.~Dennen, A.~V. Smirnov, and V.~A. Smirnov, {\it
  {Ultraviolet Properties of N=4 Supergravity at Four Loops}},  {\em Phys. Rev.
  Lett.} {\bf 111} (2013), no.~23 231302,
  [\href{http://arxiv.org/abs/1309.2498}{{\tt arXiv:1309.2498}}].

\bibitem{Bern:2014sna}
Z.~Bern, S.~Davies, and T.~Dennen, {\it {Enhanced ultraviolet cancellations in
  $\mathcal N=5$ supergravity at four loops}},  {\em Phys. Rev. D} {\bf 90}
  (2014), no.~10 105011, [\href{http://arxiv.org/abs/1409.3089}{{\tt
  arXiv:1409.3089}}].

\bibitem{Bern:2017ucb}
Z.~Bern, J.~J.~M. Carrasco, W.-M. Chen, H.~Johansson, R.~Roiban, and M.~Zeng,
  {\it {Five-loop four-point integrand of $N=8$ supergravity as a generalized
  double copy}},  {\em Phys. Rev. D} {\bf 96} (2017), no.~12 126012,
  [\href{http://arxiv.org/abs/1708.06807}{{\tt arXiv:1708.06807}}].

\bibitem{Bern:2018jmv}
Z.~Bern, J.~J. Carrasco, W.-M. Chen, A.~Edison, H.~Johansson,
  J.~Parra-Martinez, R.~Roiban, and M.~Zeng, {\it {Ultraviolet Properties of
  $\mathcal N = 8$ Supergravity at Five Loops}},  {\em Phys. Rev. D} {\bf 98}
  (2018), no.~8 086021, [\href{http://arxiv.org/abs/1804.09311}{{\tt
  arXiv:1804.09311}}].

\bibitem{Cachazo:2013hca}
F.~Cachazo, S.~He, and E.~Y. Yuan, {\it {Scattering of Massless Particles in
  Arbitrary Dimensions}},  {\em Phys. Rev. Lett.} {\bf 113} (2014), no.~17
  171601, [\href{http://arxiv.org/abs/1307.2199}{{\tt arXiv:1307.2199}}].

\bibitem{Cachazo:2013iea}
F.~Cachazo, S.~He, and E.~Y. Yuan, {\it {Scattering of Massless Particles:
  Scalars, Gluons and Gravitons}},  {\em JHEP} {\bf 07} (2014) 033,
  [\href{http://arxiv.org/abs/1309.0885}{{\tt arXiv:1309.0885}}].

\bibitem{Cachazo:2013gna}
F.~Cachazo, S.~He, and E.~Y. Yuan, {\it {Scattering equations and
  Kawai-Lewellen-Tye orthogonality}},  {\em Phys. Rev. D} {\bf 90} (2014),
  no.~6 065001, [\href{http://arxiv.org/abs/1306.6575}{{\tt arXiv:1306.6575}}].

\bibitem{Mason:2013sva}
L.~Mason and D.~Skinner, {\it {Ambitwistor strings and the scattering
  equations}},  {\em JHEP} {\bf 07} (2014) 048,
  [\href{http://arxiv.org/abs/1311.2564}{{\tt arXiv:1311.2564}}].

\bibitem{Berkovits:2013xba}
N.~Berkovits, {\it {Infinite Tension Limit of the Pure Spinor Superstring}},
  {\em JHEP} {\bf 03} (2014) 017, [\href{http://arxiv.org/abs/1311.4156}{{\tt
  arXiv:1311.4156}}].

\bibitem{Adamo:2015hoa}
T.~Adamo and E.~Casali, {\it {Scattering equations, supergravity integrands,
  and pure spinors}},  {\em JHEP} {\bf 05} (2015) 120,
  [\href{http://arxiv.org/abs/1502.06826}{{\tt arXiv:1502.06826}}].

\bibitem{Casali:2015vta}
E.~Casali, Y.~Geyer, L.~Mason, R.~Monteiro, and K.~A. Roehrig, {\it {New
  Ambitwistor String Theories}},  {\em JHEP} {\bf 11} (2015) 038,
  [\href{http://arxiv.org/abs/1506.08771}{{\tt arXiv:1506.08771}}].

\bibitem{Cachazo:2014xea}
F.~Cachazo, S.~He, and E.~Y. Yuan, {\it {Scattering Equations and Matrices:
  From Einstein To Yang-Mills, DBI and NLSM}},  {\em JHEP} {\bf 07} (2015) 149,
  [\href{http://arxiv.org/abs/1412.3479}{{\tt arXiv:1412.3479}}].

\bibitem{Cachazo:2014nsa}
F.~Cachazo, S.~He, and E.~Y. Yuan, {\it {Einstein-Yang-Mills Scattering
  Amplitudes From Scattering Equations}},  {\em JHEP} {\bf 01} (2015) 121,
  [\href{http://arxiv.org/abs/1409.8256}{{\tt arXiv:1409.8256}}].

\bibitem{He:2016mzd}
S.~He and O.~Schlotterer, {\it {New Relations for Gauge-Theory and Gravity
  Amplitudes at Loop Level}},  {\em Phys. Rev. Lett.} {\bf 118} (2017), no.~16
  161601, [\href{http://arxiv.org/abs/1612.00417}{{\tt arXiv:1612.00417}}].

\bibitem{He:2017spx}
S.~He, O.~Schlotterer, and Y.~Zhang, {\it {New BCJ representations for one-loop
  amplitudes in gauge theories and gravity}},  {\em Nucl. Phys. B} {\bf 930}
  (2018) 328--383, [\href{http://arxiv.org/abs/1706.00640}{{\tt
  arXiv:1706.00640}}].

\bibitem{Edison:2020uzf}
A.~Edison, S.~He, O.~Schlotterer, and F.~Teng, {\it {One-loop Correlators and
  BCJ Numerators from Forward Limits}},  {\em JHEP} {\bf 09} (2020) 079,
  [\href{http://arxiv.org/abs/2005.03639}{{\tt arXiv:2005.03639}}].

\bibitem{Edison:2020ehu}
A.~Edison and F.~Teng, {\it {Efficient Calculation of Crossing Symmetric BCJ
  Tree Numerators}},  {\em JHEP} {\bf 12} (2020) 138,
  [\href{http://arxiv.org/abs/2005.03638}{{\tt arXiv:2005.03638}}].

\bibitem{Adamo:2013tsa}
T.~Adamo, E.~Casali, and D.~Skinner, {\it {Ambitwistor strings and the
  scattering equations at one loop}},  {\em JHEP} {\bf 04} (2014) 104,
  [\href{http://arxiv.org/abs/1312.3828}{{\tt arXiv:1312.3828}}].

\bibitem{Geyer:2015bja}
Y.~Geyer, L.~Mason, R.~Monteiro, and P.~Tourkine, {\it {Loop Integrands for
  Scattering Amplitudes from the Riemann Sphere}},  {\em Phys. Rev. Lett.} {\bf
  115} (2015), no.~12 121603, [\href{http://arxiv.org/abs/1507.00321}{{\tt
  arXiv:1507.00321}}].

\bibitem{Geyer:2015jch}
Y.~Geyer, L.~Mason, R.~Monteiro, and P.~Tourkine, {\it {One-loop amplitudes on
  the Riemann sphere}},  {\em JHEP} {\bf 03} (2016) 114,
  [\href{http://arxiv.org/abs/1511.06315}{{\tt arXiv:1511.06315}}].

\bibitem{Geyer:2016wjx}
Y.~Geyer, L.~Mason, R.~Monteiro, and P.~Tourkine, {\it {Two-Loop Scattering
  Amplitudes from the Riemann Sphere}},  {\em Phys. Rev. D} {\bf 94} (2016),
  no.~12 125029, [\href{http://arxiv.org/abs/1607.08887}{{\tt
  arXiv:1607.08887}}].

\bibitem{Geyer:2017ela}
Y.~Geyer and R.~Monteiro, {\it {Gluons and gravitons at one loop from
  ambitwistor strings}},  {\em JHEP} {\bf 03} (2018) 068,
  [\href{http://arxiv.org/abs/1711.09923}{{\tt arXiv:1711.09923}}].

\bibitem{Geyer:2018xwu}
Y.~Geyer and R.~Monteiro, {\it {Two-Loop Scattering Amplitudes from Ambitwistor
  Strings: from Genus Two to the Nodal Riemann Sphere}},  {\em JHEP} {\bf 11}
  (2018) 008, [\href{http://arxiv.org/abs/1805.05344}{{\tt arXiv:1805.05344}}].

\bibitem{Geyer:2019hnn}
Y.~Geyer, R.~Monteiro, and R.~Stark-Much\~ao, {\it {Two-Loop Scattering
  Amplitudes: Double-Forward Limit and Colour-Kinematics Duality}},  {\em JHEP}
  {\bf 12} (2019) 049, [\href{http://arxiv.org/abs/1908.05221}{{\tt
  arXiv:1908.05221}}].

\bibitem{He:2015yua}
S.~He and E.~Y. Yuan, {\it {One-loop Scattering Equations and Amplitudes from
  Forward Limit}},  {\em Phys. Rev. D} {\bf 92} (2015), no.~10 105004,
  [\href{http://arxiv.org/abs/1508.06027}{{\tt arXiv:1508.06027}}].

\bibitem{Cachazo:2015aol}
F.~Cachazo, S.~He, and E.~Y. Yuan, {\it {One-Loop Corrections from Higher
  Dimensional Tree Amplitudes}},  {\em JHEP} {\bf 08} (2016) 008,
  [\href{http://arxiv.org/abs/1512.05001}{{\tt arXiv:1512.05001}}].

\bibitem{Mafra:2011nw}
C.~R. Mafra, O.~Schlotterer, and S.~Stieberger, {\it {Complete N-Point
  Superstring Disk Amplitude II. Amplitude and Hypergeometric Function
  Structure}},  {\em Nucl. Phys. B} {\bf 873} (2013) 461--513,
  [\href{http://arxiv.org/abs/1106.2646}{{\tt arXiv:1106.2646}}].

\bibitem{Mafra:2011nv}
C.~R. Mafra, O.~Schlotterer, and S.~Stieberger, {\it {Complete N-Point
  Superstring Disk Amplitude I. Pure Spinor Computation}},  {\em Nucl. Phys. B}
  {\bf 873} (2013) 419--460, [\href{http://arxiv.org/abs/1106.2645}{{\tt
  arXiv:1106.2645}}].

\bibitem{Bjerrum-Bohr:2016juj}
N.~E.~J. Bjerrum-Bohr, J.~L. Bourjaily, P.~H. Damgaard, and B.~Feng, {\it
  {Analytic representations of Yang\textendash{}Mills amplitudes}},  {\em Nucl.
  Phys. B} {\bf 913} (2016) 964--986,
  [\href{http://arxiv.org/abs/1605.06501}{{\tt arXiv:1605.06501}}].

\bibitem{Bjerrum-Bohr:2016axv}
N.~E.~J. Bjerrum-Bohr, J.~L. Bourjaily, P.~H. Damgaard, and B.~Feng, {\it
  {Manifesting Color-Kinematics Duality in the Scattering Equation Formalism}},
   {\em JHEP} {\bf 09} (2016) 094, [\href{http://arxiv.org/abs/1608.00006}{{\tt
  arXiv:1608.00006}}].

\bibitem{Du:2017kpo}
Y.-J. Du and F.~Teng, {\it {BCJ numerators from reduced Pfaffian}},  {\em JHEP}
  {\bf 04} (2017) 033, [\href{http://arxiv.org/abs/1703.05717}{{\tt
  arXiv:1703.05717}}].

\bibitem{Fu:2017uzt}
C.-H. Fu, Y.-J. Du, R.~Huang, and B.~Feng, {\it {Expansion of
  Einstein-Yang-Mills Amplitude}},  {\em JHEP} {\bf 09} (2017) 021,
  [\href{http://arxiv.org/abs/1702.08158}{{\tt arXiv:1702.08158}}].

\bibitem{Teng:2017tbo}
F.~Teng and B.~Feng, {\it {Expanding Einstein-Yang-Mills by Yang-Mills in CHY
  frame}},  {\em JHEP} {\bf 05} (2017) 075,
  [\href{http://arxiv.org/abs/1703.01269}{{\tt arXiv:1703.01269}}].

\bibitem{Du:2017gnh}
Y.-J. Du, B.~Feng, and F.~Teng, {\it {Expansion of All Multitrace Tree Level
  EYM Amplitudes}},  {\em JHEP} {\bf 12} (2017) 038,
  [\href{http://arxiv.org/abs/1708.04514}{{\tt arXiv:1708.04514}}].

\bibitem{Huang:2017ydz}
R.~Huang, Y.-J. Du, and B.~Feng, {\it {Understanding the Cancelation of Double
  Poles in the Pfaffian of CHY-formulism}},  {\em JHEP} {\bf 06} (2017) 133,
  [\href{http://arxiv.org/abs/1702.05840}{{\tt arXiv:1702.05840}}].

\bibitem{Cardona:2016gon}
C.~Cardona, B.~Feng, H.~Gomez, and R.~Huang, {\it {Cross-ratio Identities and
  Higher-order Poles of CHY-integrand}},  {\em JHEP} {\bf 09} (2016) 133,
  [\href{http://arxiv.org/abs/1606.00670}{{\tt arXiv:1606.00670}}].

\bibitem{DelDuca:1999rs}
V.~Del~Duca, L.~J. Dixon, and F.~Maltoni, {\it {New color decompositions for
  gauge amplitudes at tree and loop level}},  {\em Nucl. Phys. B} {\bf 571}
  (2000) 51--70, [\href{http://arxiv.org/abs/hep-ph/9910563}{{\tt
  hep-ph/9910563}}].

\bibitem{He:2018pol}
S.~He, F.~Teng, and Y.~Zhang, {\it {String amplitudes from field-theory
  amplitudes and vice versa}},  {\em Phys. Rev. Lett.} {\bf 122} (2019), no.~21
  211603, [\href{http://arxiv.org/abs/1812.03369}{{\tt arXiv:1812.03369}}].

\bibitem{He:2019drm}
S.~He, F.~Teng, and Y.~Zhang, {\it {String Correlators: Recursive Expansion,
  Integration-by-Parts and Scattering Equations}},  {\em JHEP} {\bf 09} (2019)
  085, [\href{http://arxiv.org/abs/1907.06041}{{\tt arXiv:1907.06041}}].

\bibitem{Mizera:2019blq}
S.~Mizera, {\it {Kinematic Jacobi Identity is a Residue Theorem: Geometry of
  Color-Kinematics Duality for Gauge and Gravity Amplitudes}},  {\em Phys. Rev.
  Lett.} {\bf 124} (2020), no.~14 141601,
  [\href{http://arxiv.org/abs/1912.03397}{{\tt arXiv:1912.03397}}].

\bibitem{Frost:2019fjn}
H.~Frost and L.~Mason, {\it {Lie Polynomials and a Twistorial Correspondence
  for Amplitudes}},  \href{http://arxiv.org/abs/1912.04198}{{\tt
  arXiv:1912.04198}}.

\bibitem{Gao:2017dek}
X.~Gao, S.~He, and Y.~Zhang, {\it {Labelled tree graphs, Feynman diagrams and
  disk integrals}},  {\em JHEP} {\bf 11} (2017) 144,
  [\href{http://arxiv.org/abs/1708.08701}{{\tt arXiv:1708.08701}}].

\bibitem{Arkani-Hamed:2017mur}
N.~Arkani-Hamed, Y.~Bai, S.~He, and G.~Yan, {\it {Scattering Forms and the
  Positive Geometry of Kinematics, Color and the Worldsheet}},  {\em JHEP} {\bf
  05} (2018) 096, [\href{http://arxiv.org/abs/1711.09102}{{\tt
  arXiv:1711.09102}}].

\bibitem{He:2018pue}
S.~He, G.~Yan, C.~Zhang, and Y.~Zhang, {\it {Scattering Forms, Worldsheet Forms
  and Amplitudes from Subspaces}},  {\em JHEP} {\bf 08} (2018) 040,
  [\href{http://arxiv.org/abs/1803.11302}{{\tt arXiv:1803.11302}}].

\bibitem{Feng:2020opo}
B.~Feng and Y.~Zhang, {\it {Note on the Labelled tree graphs}},  {\em JHEP}
  {\bf 12} (2020) 096, [\href{http://arxiv.org/abs/2009.02394}{{\tt
  arXiv:2009.02394}}].

\bibitem{Stieberger:2013hza}
S.~Stieberger and T.~R. Taylor, {\it {Superstring Amplitudes as a Mellin
  Transform of Supergravity}},  {\em Nucl. Phys.} {\bf B873} (2013) 65--91,
  [\href{http://arxiv.org/abs/1303.1532}{{\tt arXiv:1303.1532}}].

\bibitem{Lam:2018tgm}
C.~S. Lam, {\it {Pfaffian Diagrams for Gluon Tree Amplitudes}},  {\em Phys.
  Rev. D} {\bf 98} (2018), no.~7 076002,
  [\href{http://arxiv.org/abs/1808.07575}{{\tt arXiv:1808.07575}}].

\bibitem{Du:2016tbc}
Y.-J. Du and C.-H. Fu, {\it {Explicit BCJ numerators of nonlinear simga
  model}},  {\em JHEP} {\bf 09} (2016) 174,
  [\href{http://arxiv.org/abs/1606.05846}{{\tt arXiv:1606.05846}}].

\bibitem{Carrasco:2016ldy}
J.~J.~M. Carrasco, C.~R. Mafra, and O.~Schlotterer, {\it {Abelian Z-theory:
  NLSM amplitudes and $\alpha$'-corrections from the open string}},  {\em JHEP}
  {\bf 06} (2017) 093, [\href{http://arxiv.org/abs/1608.02569}{{\tt
  arXiv:1608.02569}}].

\bibitem{Feng:2012sy}
B.~Feng and S.~He, {\it {Graphs, determinants and gravity amplitudes}},  {\em
  JHEP} {\bf 10} (2012) 121, [\href{http://arxiv.org/abs/1207.3220}{{\tt
  arXiv:1207.3220}}].

\bibitem{Arkani-Hamed:2023jry}
N.~Arkani-Hamed, Q.~Cao, J.~Dong, C.~Figueiredo, and S.~He, {\it
  {Scalar-Scaffolded Gluons and the Combinatorial Origins of Yang-Mills
  Theory}},  \href{http://arxiv.org/abs/2401.00041}{{\tt arXiv:2401.00041}}.

\bibitem{Arkani-Hamed:2023swr}
N.~Arkani-Hamed, Q.~Cao, J.~Dong, C.~Figueiredo, and S.~He, {\it {Hidden zeros
  for particle/string amplitudes and the unity of colored scalars, pions and
  gluons}},  \href{http://arxiv.org/abs/2312.16282}{{\tt arXiv:2312.16282}}.

\bibitem{Chen:2019ywi}
G.~Chen, H.~Johansson, F.~Teng, and T.~Wang, {\it {On the kinematic algebra for
  BCJ numerators beyond the MHV sector}},  {\em JHEP} {\bf 11} (2019) 055,
  [\href{http://arxiv.org/abs/1906.10683}{{\tt arXiv:1906.10683}}].

\bibitem{Cheung:2016prv}
C.~Cheung and C.-H. Shen, {\it {Symmetry for Flavor-Kinematics Duality from an
  Action}},  {\em Phys. Rev. Lett.} {\bf 118} (2017), no.~12 121601,
  [\href{http://arxiv.org/abs/1612.00868}{{\tt arXiv:1612.00868}}].

\bibitem{Arkani-Hamed:2016rak}
N.~Arkani-Hamed, L.~Rodina, and J.~Trnka, {\it {Locality and Unitarity of
  Scattering Amplitudes from Singularities and Gauge Invariance}},  {\em Phys.
  Rev. Lett.} {\bf 120} (2018), no.~23 231602,
  [\href{http://arxiv.org/abs/1612.02797}{{\tt arXiv:1612.02797}}].

\bibitem{Rodina:2020jlw}
L.~Rodina, {\it {UV consistency conditions for Cachazo-He-Yuan integrands}},
  {\em Phys. Rev. D} {\bf 102} (2020), no.~4 045012,
  [\href{http://arxiv.org/abs/2005.06446}{{\tt arXiv:2005.06446}}].

\bibitem{Frost:2020eoa}
H.~Frost, C.~R. Mafra, and L.~Mason, {\it {A Lie bracket for the momentum
  kernel}},  \href{http://arxiv.org/abs/2012.00519}{{\tt arXiv:2012.00519}}.

\bibitem{Mafra:2020qst}
C.~R. Mafra, {\it {Planar binary trees in scattering amplitudes}},  11, 2020.
\newblock \href{http://arxiv.org/abs/2011.14413}{{\tt arXiv:2011.14413}}.

\bibitem{Bridges:2021ebs}
E.~Bridges and C.~R. Mafra, {\it {Local BCJ numerators for ten-dimensional SYM
  at one loop}},  {\em JHEP} {\bf 07} (2021) 031,
  [\href{http://arxiv.org/abs/2102.12943}{{\tt arXiv:2102.12943}}].

\bibitem{Agerskov:2019ryp}
J.~Agerskov, N.~E.~J. Bjerrum-Bohr, H.~Gomez, and C.~Lopez-Arcos, {\it
  {One-Loop Yang-Mills Integrands from Scattering Equations}},  {\em Phys. Rev.
  D} {\bf 102} (2020), no.~4 045023,
  [\href{http://arxiv.org/abs/1910.03602}{{\tt arXiv:1910.03602}}].

\bibitem{Mafra:2016ltu}
C.~R. Mafra, {\it {Berends-Giele recursion for double-color-ordered
  amplitudes}},  {\em JHEP} {\bf 07} (2016) 080,
  [\href{http://arxiv.org/abs/1603.09731}{{\tt arXiv:1603.09731}}].

\end{thebibliography}\endgroup

\end{document}